\documentclass[trackchanges]{aastex701}
\usepackage{amsmath}
\usepackage{graphicx}
\usepackage{subcaption}
\usepackage{adjustbox}

\definecolor{green1}{rgb}{0.2, 0.6, 0.1}
\definecolor{purp1}{rgb}{0.6, 0.1, 0.45}

\newcommand{\ALMANAC}{\texttt{ALMANAC}}

\def\d{{\rm d}}
\newcommand{\rmd}{\mathrm{d}}

\def\d{\mathrm{d}}

\def\Bv{{\boldsymbol B}}

\begin{document}
    
    \title{Multi-Thermal CME Detection with \ALMANAC}
    \correspondingauthor{Thomas Williams}
    \email{tomwilliamsphd@gmail.com}
        
    \author[0000-0002-2006-6096]{Thomas Williams}
    \email{tomwilliamsphd@gmail.com}
    \affiliation{Department of Mathematical Sciences, Durham University, Durham, UK}
    
    \author[0000-0003-4015-5106]{Christopher B. Prior}
    \email{christopher.b.prior@durham.ac.uk}
    \affiliation{Department of Mathematical Sciences, Durham University, Durham, UK}
    
    \author[0000-0003-2297-9312]{David MacTaggart}
    \email{david.mactaggart@glasgow.ac.uk}
    \affiliation{School of Mathematics \& Statistics, University of Glasgow, Glasgow, UK}
    
    \author[0000-0002-6547-5838]{Huw Morgan}
    \email{hum@aber.ac.uk}
    \affiliation{Department of Physics, Aberystwyth University, Aberystwyth, UK}

\begin{abstract}
Reliable identification of low-coronal CME origins remains a key limitation in space weather forecasting with coronagraphs not directly resolving low-coronal signatures. We present a re-engineered multi-thermal implementation of the Automated Detection of CoronaL MAss Ejecta origiNs for Space Weather AppliCations (\texttt{ALMANAC}) algorithm, designed to detect eruptive signatures in EUV observations from SDO/AIA. The framework extends the original method to a multi-wavelength system, improving robustness against projection effects, instrumental artefacts, and wavelength-dependent ambiguities via complementary temperature responses. A spatio-temporal clustering scheme merges detections across channels, reducing event bifurcation and improving coherence while maintaining near-real-time performance through parallel computing. Benchmarking against 20 halo CMEs from \texttt{CDAW} shows improved interpretability and operational usability, with clearer separation of eruptions and more consistent onset localisation relative to coronagraph estimates. The main benefits arise from improved event discrimination, reduced fragmentation, and more interpretable source region identification. \texttt{ALMANAC} shows sensitivity to precursor low-coronal activity not always captured in white-light catalogues, highlighting the advantages of EUV-based detection for early warning. When coupled with the Active Region Topology framework, it enables co-analysis of coronal intensity variability and photospheric magnetic evolution. In this context, kurtosis time-series from multi-wavelength EUV data exhibit recurrent pre-eruptive spikes that frequently align with enhancements in magnetic winding and helicity injection. Across multiple regions, these signatures often precede solar activity, including potential discrimination of X-class flare events, while remaining suppressed during magnetically quiet intervals. Overall, integrating coronal statistical diagnostics with photospheric topology offers a pathway toward improved eruption forecasting and more physically grounded space weather prediction.
\end{abstract}

\keywords{\uat{Space Weather}{2037} --- \uat{Solar Activity}{1475} --- \uat{Solar Coronal Mass Ejections}{310} --- \uat{Solar Flares}{1496} --- \uat{Solar Physics}{1476}}

\section{Introduction}\label{sec:intro}
Space weather encompasses the dynamic solar and heliospheric processes that perturb the near-Earth environment and influence both technological systems and human activities. The primary drivers of severe space weather are solar flares and coronal mass ejections (CMEs), which release vast quantities of electromagnetic radiation, energetic particles, and magnetised plasma into interplanetary space. Solar flares can generate intense ionospheric disturbances that disrupt high-frequency radio communication, satellite navigation systems, and radar operations, while CMEs are the principal cause of major geomagnetic storms at Earth \citep{schwenn2006,riley2018}. These disturbances can damage spacecraft electronics, increase atmospheric drag on low-Earth-orbit satellites, and produce geomagnetically induced currents capable of degrading or destroying power-grid infrastructure \citep{horne2013,johnson2016}. In addition, solar energetic particle events associated with eruptive solar activity pose substantial radiation hazards to astronauts and avionics systems, particularly for missions beyond the protection of the terrestrial magnetosphere \citep{klein2018}. As modern society becomes increasingly dependent on space-based technologies and interconnected electrical systems, reliable forecasting of extreme solar activity has become a critical objective of contemporary heliophysics and space weather research.

Considerable effort has therefore been directed toward forecasting solar flares using both physics-based and data-driven approaches. Early statistical forecasting methods relied heavily on sunspot classifications and magnetic complexity measures such as the McIntosh and Mount Wilson schemes, which demonstrated moderate skill in identifying flare-productive active regions \citep{mcintosh1990,bloomfield2012}. More recently, machine learning techniques using vector magnetogram data from instruments such as the Helioseismic and Magnetic Imager onboard the Solar Dynamics Observatory have significantly improved predictive capabilities by exploiting large parameter spaces and nonlinear relationships within active region magnetic fields \citep{bobra2015,nishizuka2018}. Parameters associated with magnetic free energy, shear, electric currents, and helicity have proven particularly valuable for distinguishing eruptive from non-eruptive active regions \citep{leka2003,toriumi2019}. Nevertheless, substantial challenges remain. Flare occurrence is intrinsically stochastic, positive training samples are relatively rare compared with non-flaring intervals, and many algorithms suffer from poor interpretability and limited generalisation across solar cycles. Furthermore, although modern machine learning methods often achieve high skill scores, operational forecasting performance is frequently reduced by class imbalance, differing validation procedures, and limited physical understanding of the triggering mechanisms underlying flare initiation \citep{barnes2016,florios2018}.

Recent work has established magnetic winding as a promising topological diagnostic for understanding the build-up and release of magnetic energy in solar active regions. Unlike magnetic helicity, which is weighted by magnetic flux strength, magnetic winding isolates the geometric entanglement and braiding of magnetic field lines, providing a more direct measure of magnetic topology within emerging and evolving active regions \citep{prior2019,prior2020}. Through a combination of theoretical modelling and observational analysis using vector magnetograms, this work has demonstrated that winding can identify signatures of twisted magnetic flux-tube emergence and retain information about the subsurface magnetic structure responsible for active region formation \citep{mactaggart2021}. Subsequent studies showed that flare-productive active regions exhibit enhanced and rapidly evolving winding flux signatures compared with flare-quiet regions, suggesting that winding captures the accumulation of non-potential magnetic topology associated with eruptive activity \citep{raphaldini2022}. The work of \citet{williams25,williams26} builds upon this by incorporating topologically derived winding and helicity time series into machine-learning flare forecasting frameworks, where they provide physically interpretable predictors that may improve the early identification of eruptive solar activity. Together, these studies demonstrate that magnetic winding offers a valuable bridge between mathematically rigorous descriptions of solar magnetic topology and operational space weather forecasting applications.

In parallel with flare forecasting, extensive research has focused on CME detection, tracking, and prediction of geoeffectiveness at Earth. Automated CME catalogues derived from coronagraph observations, including the Coordinated Data Analysis Workshops catalogue \citep[\texttt{CDAW};][]{gopalswamy2009catalog,gopalswamy2009halo}, Computer Aided CME Tracking system \citep[\texttt{CACTus};][]{robbrecht2004cactus,robbrecht2009automated}, and Coronal Image Processing catalogue \citep[\texttt{CORIMP};][]{morgan2006multiscale,byrne2012corimp}, have enabled large-scale statistical studies of CME properties and improved event identification consistency \citep{yashiro2004,robbrecht2009automated,byrne2012corimp}. Heliospheric imaging and magnetohydrodynamic propagation models such as \texttt{ENLIL} \citep{odstrcil2003,mays2015} and Heliospheric Upwind eXtrapolation with time dependence model \citep[\texttt{HUXt};][]{owens2020,barnard2022} are now routinely used to estimate CME arrival times and geomagnetic impact probabilities \citep{odstrcil2003,mays2015}. These methods have achieved typical CME arrival-time errors of approximately 10 to 15 hours, representing substantial progress for operational space weather forecasting. However, significant uncertainties remain because CME evolution depends strongly on interactions with the ambient solar wind, CME-CME interactions, and the poorly constrained internal magnetic structure of eruptions \citep{temmer2021}. In particular, predicting the southward magnetic field component responsible for strong geomagnetic storms remains one of the central unresolved problems in space weather forecasting. Furthermore, stealth CMEs and projection effects continue to complicate reliable identification and tracking from single-viewpoint coronagraph observations \citep{howard2013}.

More recent studies have increasingly incorporated machine-learning and deep-learning methodologies into CME detection pipelines in an effort to improve robustness, sensitivity to faint structures, and early-event identification. \citet{wang2019camel} developed the CME Automatic detection and tracking with MachinE Learning (\texttt{CAMEL}) framework, which combines convolutional neural networks, graph-cut segmentation, and temporal tracking algorithms to identify and characterise CMEs in coronagraph data. \texttt{CAMEL} demonstrated improved sensitivity to weak and diffuse CME structures relative to several earlier automated catalogues and was capable of detecting eruptions at earlier stages of propagation. Whereas \citet{liu2026} introduced few-shot learning approaches for CME detection, motivated by the limited availability of labelled CME training data and the need for scalable automated cataloguing during periods of high solar activity. These approaches aim to generalise CME recognition capabilities from relatively small annotated datasets, reducing dependence on extensive manual labelling while maintaining competitive detection accuracy. Parallel developments have also focused on integrating CME imagery, solar wind parameters, and machine-learning ensemble methods for CME arrival-time forecasting at Earth. The \texttt{CMETNet} framework \citep{cmetnet} combines coronagraph image analysis with numerical and solar wind data using ensemble learning techniques, achieving CME arrival-time prediction errors below approximately 10 hours, comparable to or better than many operational physics-based models.

Despite these advances, substantial challenges remain for operational CME forecasting. Many machine-learning methods require carefully curated training datasets, often struggle to generalise across instruments or solar cycles, and may lack the physical interpretability associated with physics-based heliospheric models. In addition, even highly accurate coronagraph detections cannot directly determine the internal magnetic structure of CMEs, which remains one of the principal unresolved limitations in forecasting geomagnetic storm intensity at Earth \citep{temmer2021}. \citet{gandhi2024,gandhi2025} focused on improving the three-dimensional detection and characterisation of coronal mass ejections using both geometrical modelling and machine-learning techniques. Their studies demonstrated that correcting projection effects through multi-viewpoint Graduated Cylindrical Shell (GCS) modelling substantially improves estimates of CME speed, width, and propagation direction, while more recent deep-learning approaches using convolutional neural networks have shown promise for automated prediction of CME kinematic parameters directly from coronagraph imagery.

A common pitfall in all these methods mentioned is that they do not natively provide information regarding the low-coronal source region of eruptions, with reliable observational coverage generally beginning only between approximately 2 and 6 solar radii above the visible photosphere. In contrast, the Automated Detection of CoronaL MAss Ejecta origiNs for Space Weather AppliCations \citep[\texttt{ALMANAC};][]{almanac} operates directly on EUV observations from the Atmospheric Imaging Assembly \citep[AIA;][]{lemen2012aia} aboard the Solar Dynamics Observatory \citep[SDO;][]{pesnell2012sdo}, enabling the detection of eruptive activity in the low corona before the CME becomes visible in coronagraph fields of view. By identifying source regions and eruption onset times at earlier stages, \texttt{ALMANAC} provides information that may improve the initialization of CME propagation models and enhance real-time space weather alert capabilities. The method was developed to address the need for rapid and reliable CME detection for operational space weather forecasting, particularly for Earth-directed events where early warnings are essential. 

The original, single-channel implementation of \texttt{ALMANAC} detects eruptive EUV signatures using image normalization, high-bandpass filtering, and running ratio or difference imaging to isolate transient coronal features. Candidate eruption regions are identified through spatio-temporal thresholding and clustered across consecutive frames to estimate eruption onset times and heliographic source locations. Validation against 20 halo CMEs from the \texttt{CDAW} catalogue during Solar Cycle 24 showed good agreement, with mean positional deviations of approximately $11^{\circ}$ and temporal offsets of about 40\,minutes. A further study by \citet{aslam24} investigated 38 CME events and compared magnetic winding signatures to their origins as determined by \texttt{ALMANAC}. Their findings showed close agreement between the photospheric magnetic winding calculations and \texttt{ALMANAC} for all bar one of the 38 CMEs where \texttt{ALMANAC} failed to detect the eruption. This illustrates that the method, while promising, requires improvement as the single-channel method remains limited by projection effects, threshold sensitivity, and spurious detections caused by instrumental artefacts and non-eruptive coronal variability. In this manuscript, we present an updated multi-thermal implementation of \texttt{ALMANAC} designed to mitigate these limitations while improving the interpretability and operational applicability of the method. Furthermore, we also extend the \texttt{ALMANAC} framework to distinguish between CMEs and flares, and explore the efficacy of the method being used to not only detect space weather events but to also predict them.

\section{Numerical Methods}\label{sec:method}
We present a full Python re-implementation of the \ALMANAC\ algorithm originally developed in IDL. The pipeline is designed to automatically identify eruptive events in SDO/AIA extreme ultraviolet (EUV) observations, determine their onset times, and estimate their low-coronal source locations. In contrast to the original implementation, this version emphasizes modularity, scalability, and robustness, enabling high-throughput processing of multi-wavelength datasets using modern Python scientific libraries.

\begin{figure*}
\centerline{\includegraphics[width=0.925\textwidth]{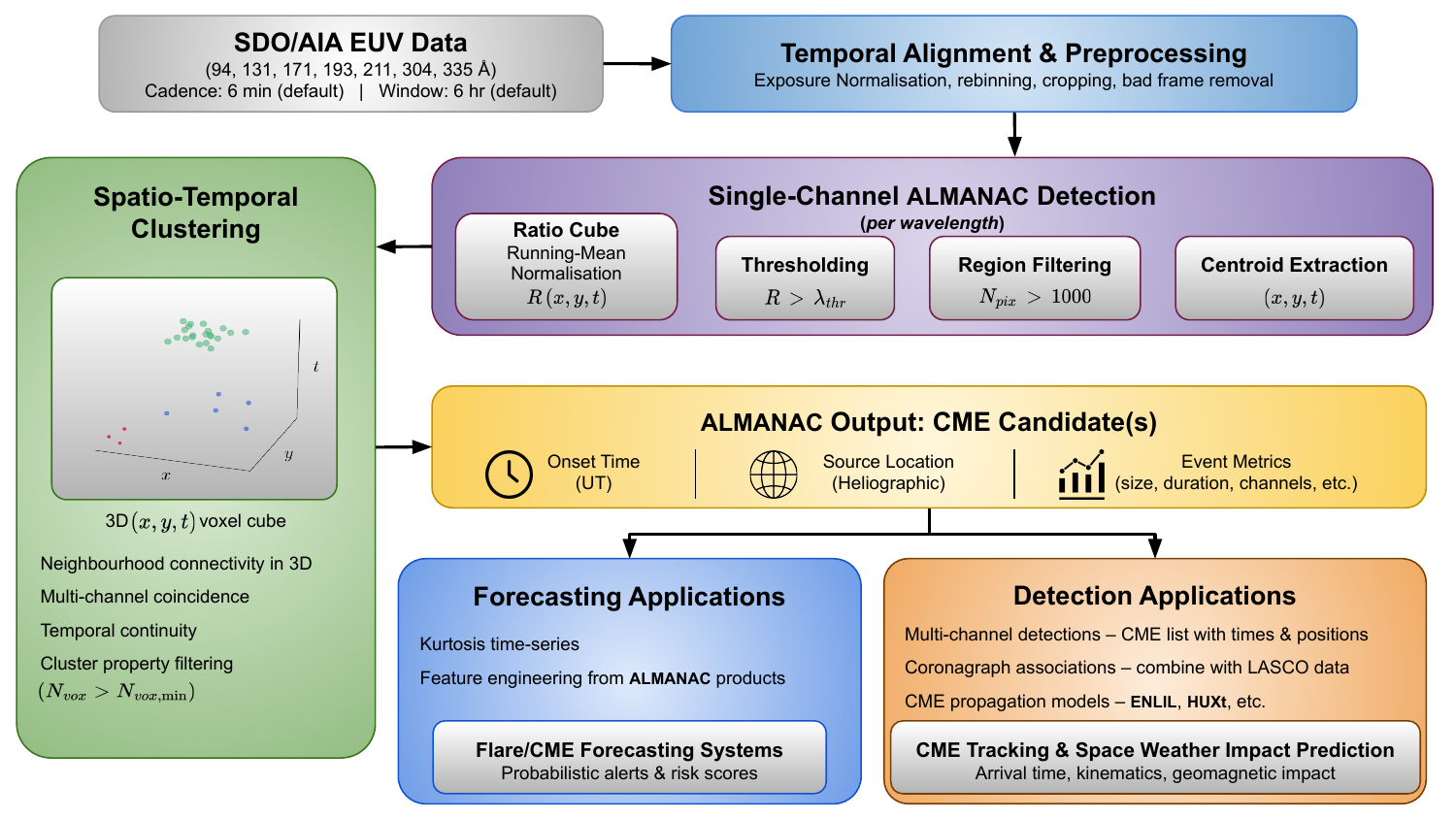}}
\caption{Flowchart illustrating the five stages of the \texttt{ALMANAC} algorithm with two example use-cases for the outputs with respect to space weather forecasting.}
\label{fig:flow}
\end{figure*}

The workflow (Figure\,\ref{fig:flow}) proceeds through five primary stages: (1) acquisition and temporal alignment of AIA synoptic data, (2) wavelength-specific detection of candidate eruptive regions using spatio-temporal intensity analysis, (3) centroid extraction and heliographic coordinate transformation, (4) construction of a three-dimensional spatio-temporal clustering cube to associate multi-wavelength detections, and (5) identification and validation of CME source regions. The pipeline operates either on a per-event basis or across extended time windows, supporting both targeted CME analysis and survey-style studies.

Due to \texttt{ALMANAC} being a stand-alone module, it is well-suited to complement other forecasting methods. As such, we also briefly outline the Active Region Topology \citep[\texttt{ARTop};][]{alielden23} code which is used to further illustrate the forecasting efficacy of \texttt{ALMANAC} in \S\ref{sec:kurt} through the use of kurtosis time-series.

\subsection{Data Acquisition and Temporal Standardisation}\label{sec:data_acq}
For a given target end time, AIA EUV images are retrieved over a configurable temporal window (typically 6–8 hours) using synoptic or near-real-time data products. Data are downloaded at a fixed cadence (default 6 minutes), ensuring uniform temporal sampling across wavelengths. Prior to processing, the requested end time is snapped to the nearest valid cadence boundary to maintain strict alignment between channels. This end time can be user specified or automated from other detection suites such as LASCO/C2 CME \citep{domingo1995soho,brueckner1995lasco} identifications.

Images are grouped by wavelength and processed independently in parallel, with each wavelength handled by a separate process. This design mirrors the original \ALMANAC\ philosophy while leveraging Python’s multiprocessing framework to scale efficiently across available CPU cores. Each wavelength channel produces an independent set of candidate eruptive regions, which are later merged through clustering.

\subsection{Image Preprocessing}\label{sec:preproc}
For each wavelength, all available AIA FITS files within the analysis window are read into a three-dimensional data cube with dimensions ($x$, $y$, $t$). Frames with invalid or extreme median intensity values are removed using a robust median-based filter, whilst the remaining frames are normalised by exposure time and scaled to a consistent median intensity level to minimise inter-frame variability unrelated to solar activity. More detail on this normalisation can be found in \citet{almanac}.

Spatial re-binning is applied when necessary to limit computational cost and standardize image dimensions. Images larger than 1\,MP are block-averaged to this resolution, i.e. when science grade data is utilised which is 16\,MP. The field of view is further cropped to exclude peripheral regions near the poles and off-disk as the method presented here focuses on potential Earth-directed Halo CMEs. The resulting resolution of $666\times870$ pixels is 55\,\% the size of the synoptic/NRT data, with a non-square shape so as to maintain more of the information from lower latitudes.

Eruptive events are identified through temporal variability rather than absolute brightness. For each pixel, a running temporal mean is computed using a reflective padding scheme to avoid edge artifacts. The absolute deviation from this running mean is calculated, producing a difference cube that highlights transient activity. A per-pixel median of this deviation over time is used as a baseline, yielding a normalised ratio cube:
\begin{equation}
R(x,y,t)
=
\frac{
\left| I(x,y,t) - \left\langle I(x,y,t) \right\rangle_{\Delta t} \right|
}{
\mathrm{median}_{t}
\left(
\left| I(x,y,t) - \left\langle I(x,y,t) \right\rangle_{\Delta t} \right|
\right)
}
\end{equation}
where $I(x,y,t)$ is the exposure-normalised EUV intensity at pixel $\left(x,y\right)$ and time $t$. $\left\langle I \right\rangle_{\Delta t}$ is the temporal running mean computed over a symmetric window $\Delta t$, and $\mathrm{median}_{t}$ is the median evaluated over the full time axis at each pixel. 

The use of a per-pixel temporal median in the denominator normalises the ratio to the typical level of variability at each location, suppressing static coronal structure while enhancing transient eruptive signatures. The resultant ratio cube is bound to the range $0 \le R \le 30$ to suppress extreme outliers arising from low-variance background pixels and to further stabilise subsequent spatial filtering and region detection.

\subsection{Event Characterisation}\label{sec:detection}
Candidate eruptive regions are identified through a multi-stage masking process applied to the ratio cube. For each time frame, pixels exceeding a wavelength-dependent threshold are grouped using a region-size filter that removes small, isolated features unlikely to correspond to CME activity. Here, the thresholds are set to clusters of 1000 pixels exceeding $R = 2$ per wavelength.

The resulting Boolean mask is spatially smoothed using Gaussian kernels and temporally smoothed with a narrower kernel to enforce continuity in time. A second region-size filter is applied to the smoothed mask to remove fragmented or transient detections. Connected-component labelling is then performed on the three-dimensional mask volume, yielding candidate regions with distinct spatial and temporal extents.

Regions with voxel counts below a minimum volume threshold of 6000 (compared to 9000 in \citealp{almanac}) are discarded. This criterion enforces persistence and spatial coherence, reducing contamination from noise, wave phenomena, or localised brightenings unrelated to potential CMEs. These steps relax the empirically determined thresholds inherited from the original \ALMANAC\ implementation, which has a tendency to miss some smaller CME events \citep{aslam24}.

For each surviving candidate region, the temporal span of the event is evaluated to ensure a minimum duration (18 minutes, 3 frames) consistent with eruptive activity. Within each time step of the region, the centre of mass is computed in pixel space using the ratio values as weights. This weighting emphasizes the most dynamically evolving parts of the region, providing a physically motivated estimate of the eruption’s origin.

Pixel-space centroids are transformed into heliographic Stonyhurst and Carrington coordinates using \texttt{SunPy}'s World Coordinate System utilities. This transformation accounts for spacecraft pointing, plate scale, and solar ephemerides. The centroid positions, along with associated metadata, are stored for each time step, producing a time series of candidate source locations for the event and perceived direction of travel across the solar disk.

Each detected event is serialised into a compressed Python \texttt{pickle} file containing the image cutouts, masks, ratio maps, centroid coordinates, and header information. These files form the fundamental analysis units used in later clustering stages. In parallel, higher-order statistical diagnostics, including kurtosis of the intensity and ratio distributions (discussed later), are computed and saved for potential post-processing analyses. As with the single-channel method, optional visualisation products, including multi-frame movies combining EUV intensity and detection masks, can be generated for manual inspection and validation. Full details on the single-channel method and thresholding can be found in \citet{almanac}.

\subsection{Spatio-Temporal Clustering}\label{sec:cluster}
To associate detections across wavelengths and isolate physically coherent CME events, all single-channel detection products generated within the analysis window are aggregated into a three-dimensional spatio-temporal data cube. Each detection contributes a set of centroids corresponding to individual time steps during which the region is present. These centroids are defined in pixel coordinates on the cropped AIA image grid and are paired with their corresponding observation times.

Spatial binning is performed independently along the $x$ and $y$ axes using fixed-width bins (default size of 16 pixels), chosen to be large enough to accommodate small centroid shifts arising from projection effects, wavelength-dependent morphology, and temporal evolution of the eruption, while remaining small compared to the overall size of active regions. Temporal binning is applied using uniform bins corresponding to a fixed cadence (default 36 minutes), which is intentionally broader than the native \ALMANAC\ cadence in order to group detections that are closely spaced in time but may not occur in exactly the same frame across all wavelengths.

Each detection centroid is mapped to a single voxel in this $\left(x, y, t\right)$ grid based on its binned spatial location and time index. When multiple detections from different wavelengths fall into the same voxel, the voxel value is incremented accordingly. The resulting cube therefore encodes both the spatio-temporal density of detections and the degree of multi-wavelength coincidence, with higher voxel values indicating repeated or multi-channel detections at the same location and time.

Following construction of the clustering data cube, three-dimensional connected-component labelling is applied using face-connected (6-connected) neighbourhood criteria. This step identifies contiguous clusters of occupied voxels in combined spatial and temporal space, effectively grouping detections that persist across multiple adjacent time bins and neighbouring spatial bins. Clusters with a total voxel count below a specified minimum threshold $N_{\mathrm{vox,min}}=4$ are discarded, as such sparse clusters are unlikely to represent physically meaningful eruptive events and are more consistent with isolated noise, transient brightenings, or wavelength-specific artifacts from our testing.

This clustering strategy exploits the expectation that genuine CME-related activity will produce coherent, sustained signatures that are observable in multiple EUV channels and over successive time steps, whereas spurious detections tend to be localised, short-lived, and channel-dependent. By requiring spatio-temporal continuity and multi-wavelength coincidence, the clustering stage serves as a critical filter that converts channel-level detections into a small number of robust, physically plausible CME candidates.

\subsection{CME Origin Determination}\label{sec:origin}
For each validated spatio-temporal cluster, heliographic source coordinates are derived from the ensemble of contributing detections. Specifically, all Stonyhurst longitude and latitude values associated with single-channel centroids belonging to the cluster are collected across wavelengths and time steps. This ensemble may contain outliers arising from the transient misidentification of centroids, wavelength-dependent emission morphology, or increased projection effects near the solar limb.

To mitigate the influence of such outliers, robust statistical filtering is applied independently to the longitude and latitude distributions. For each coordinate, the first ($Q_1$) and third ($Q_3$) quartiles are computed, and the interquartile range is defined as $IQR = Q_3 - Q_1$. Coordinate values lying outside the interval
\begin{equation}
    \left[ Q_1 - 1.5\,IQR,\,\,Q_3 + 1.5\,IQR\right],
\end{equation}
are rejected. This criterion suppresses extreme deviations while preserving the core of the distribution, and does not assume an underlying Gaussian error model.

Following outlier rejection, the final CME origin is defined as the mean of the remaining longitude and latitude values. This averaging is performed separately for each coordinate and yields a single representative heliographic position for the event. The mean is preferred over the median at this stage because the IQR filtering removes pathological values, allowing the mean to retain sensitivity to subtle but systematic shifts in centroid position that may reflect genuine evolution of the eruptive structure.

The resulting source location therefore represents a robust, multi-wavelength, time-integrated estimate of the low-coronal origin of the eruption. By combining detections across multiple EUV channels and time steps while rejecting anomalous measurements, this approach reduces sensitivity to individual false detections that arise due to noise, partial occultation, or wavelength-specific emission features.

In \citet{almanac}, the onset time is taken as the first frame in which a detection is present. Here, this largely remains the same, except this is now the first frame for which a valid voxel exists in the clustering data cube. This, combined with the on-disk location provide the \ALMANAC\ origin for potential CME detections.

\subsection{ARTop Synergy}\label{sec:ARTop}
Whilst \texttt{ALMANAC} is a standalone module, its real strength comes from the ability to merge with other forecasting suites to provide additional or complimentary data such as from coronagraph detection algorithms or flare forecasting metrics. For example, \citet{alielden23} introduced the open-source Active Region Topology (\texttt{ARTop}) package for estimating magnetic helicity and magnetic winding injection from vector magnetogram observations. \texttt{ARTop} uses Space-weather HMI Active Region Patch \citep[SHARP;][]{hoeksema14} vector magnetic field data from the Helioseismic and Magnetic Imager \citep[HMI;][]{hmipaper} aboard SDO. Photospheric velocity fields are inferred using the Differential Affine Velocity Estimator for Vector Magnetograms (\texttt{DAVE4VM}; \citealp{schuck08}), which estimates the apparent motion of magnetic field line footpoints under the assumption of ideal evolution. From these velocities, \texttt{ARTop} computes the relative rotational rate between pairs of magnetic footpoints across the photospheric boundary, allowing for the calculation of both magnetic winding and magnetic helicity flux densities \citep{prior2020magnetic}. The helicity flux corresponds to the flux-weighted topological entanglement of the magnetic field, while the winding flux measures the same geometric complexity without magnetic flux weighting, providing complementary sensitivity to strongly sheared or emerging structures near polarity inversion lines.

A key feature of \texttt{ARTop} is the decomposition of the observed magnetic field into potential and current-carrying components using a Helmholtz decomposition,
\begin{equation}
\Bv = \Bv_p + \Bv_c,
\end{equation}
where the potential field $\Bv_p$ is uniquely determined from the observed normal magnetic field component at the photosphere, and $\Bv_c$ represents the non-potential, current-carrying contribution. Separate helicity and winding flux densities are then calculated for both components, yielding $\mathcal{H}_p'$, $\mathcal{H}_c'$, $\mathcal{L}_p'$, and $\mathcal{L}_c'$. Since eruptive activity is expected to be associated with increasing non-potential structure, \texttt{ARTop} additionally evaluates the imbalance between current-carrying and potential contributions through the quantities
\begin{equation}\label{eq:dL}
\delta L' = \int_P \left( \left| \mathcal{L}_c' \right| - \left| \mathcal{L}_p' \right| \right)\,\rmd^2x,
\end{equation}
and
\begin{equation}\label{eq:dH}
\delta H' = \int_P \left( \left| \mathcal{H}_c' \right| - \left| \mathcal{H}_p' \right| \right)\,\rmd^2x.
\end{equation}
Positive values indicate dominance of current-carrying topology over the potential background field. Previous studies have shown these current-carrying imbalance measures to be more robust to the velocity-smoothing parameter required by \texttt{DAVE4VM} and to exhibit stronger and more consistent relationships with flare and coronal mass ejection activity than the total helicity or winding fluxes alone \citep{brenopaper,raphaldini2023deciphering,aslam24}. As such, \citet{williams25} deconvolved the expressions further into positive and negative components whereby positive (negative) would indicate that the dominant input of topology at the photospheric level is current-carrying (potential) topology. Subsequently, the conditional expressions are:
\begin{equation}\label{eq:dLc_dHc}
   \delta\mathcal{L}_c^{\prime} =
    \begin{cases}
       \delta\mathcal{L}^{\prime}, & \text{if } \delta\mathcal{L}^{\prime} > 0 \\
        0,                                       & \text{otherwise}
\end{cases}
\quad\text{,}\quad
    \delta\mathcal{H}_c^{\prime} =
    \begin{cases}
       \delta\mathcal{H}^{\prime}, & \text{if } \delta\mathcal{H}^{\prime} > 0 \\
        0,                                       & \text{otherwise}
\end{cases}
,
\end{equation}
and
\begin{equation}\label{eq:dLp_dHp}
    \delta\mathcal{L}_p^{\prime} =
    \begin{cases}
        \delta\mathcal{L}^{\prime}, & \text{if } \delta\mathcal{L}^{\prime} < 0 \\
        0,                                       & \text{otherwise}
\end{cases}
\quad\text{,}\quad
    \delta\mathcal{H}_p^{\prime} =
    \begin{cases}
        \delta\mathcal{H}^{\prime}, & \text{if } \delta\mathcal{H}^{\prime} < 0 \\
        0,                                       & \text{otherwise}
\end{cases}
.
\end{equation}
For this manuscript, it is the current-carrying components $\delta\mathcal{L}_c^{\prime}$ and $\delta\mathcal{H}_c^{\prime}$ that we will focus upon. Full details on the code can be found in \citet{alielden23}, whilst \citet{williams25,williams26} outline its efficacy for flaring and use in machine learning (ML) algorithms.

\section{Results and Analysis}\label{sec:res}
\begin{figure*}
\centerline{\includegraphics[width=\textwidth]{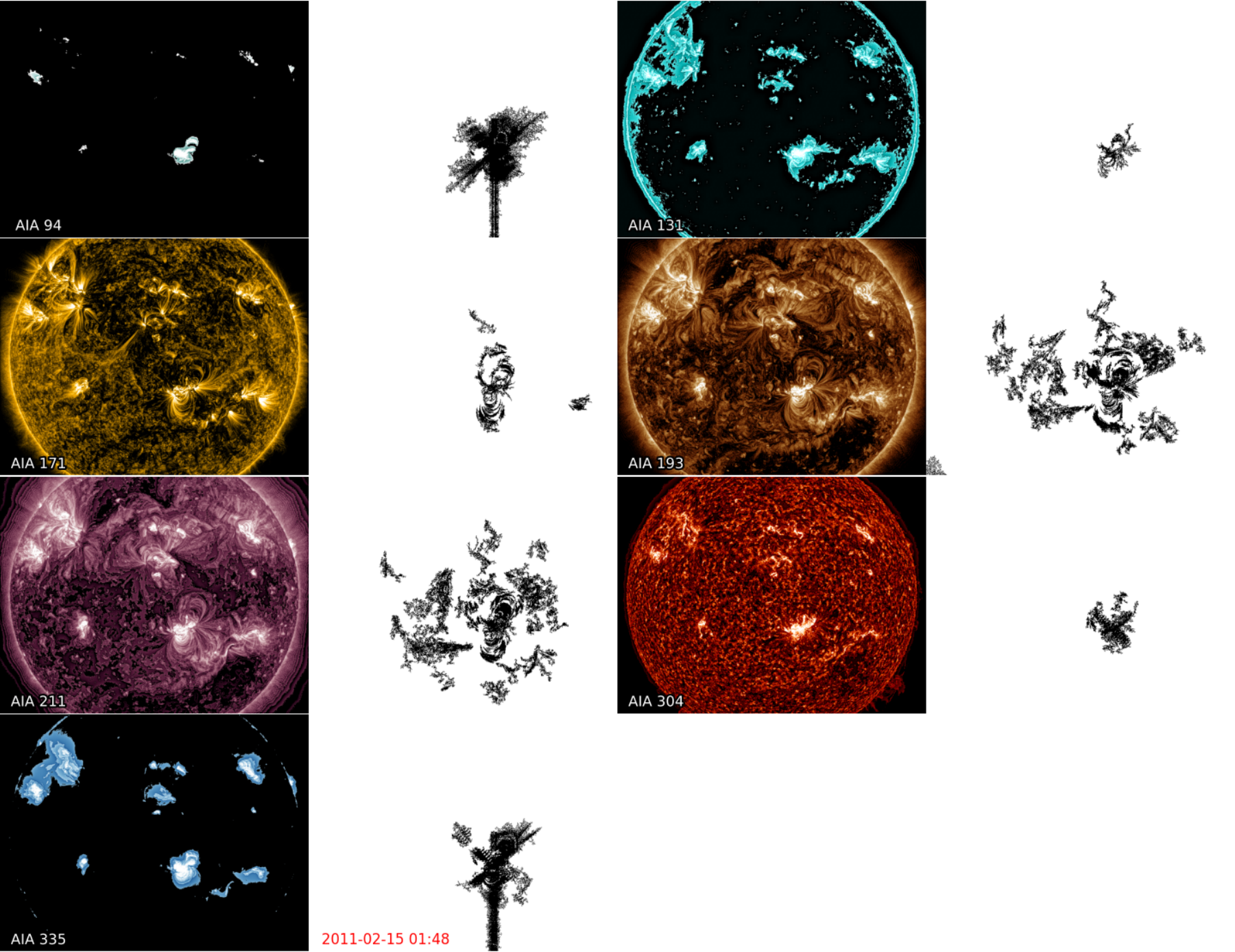}}
\caption{Example CME and associated X2.2 solar flare from AR\,11158 (\texttt{CDAW} Index 2) captured in \texttt{ALMANAC} across all SDO/AIA channels (\texttt{ALMANAC} Index 3). On the left of each panel are the SDO/AIA emission profiles that have been sharpened with Multi-Scale Gaussian Normalisation \citep[MGN]{morgan14} and the corresponding \texttt{ALMANAC} detection mask (black) on the right. Note that an animated version is only available in the online version.}
\label{fig:cme2}
\end{figure*}
\begin{figure}[ht]
    \centering

    \begin{subfigure}{0.7\textwidth}
        \centering
        \includegraphics[width=\linewidth]{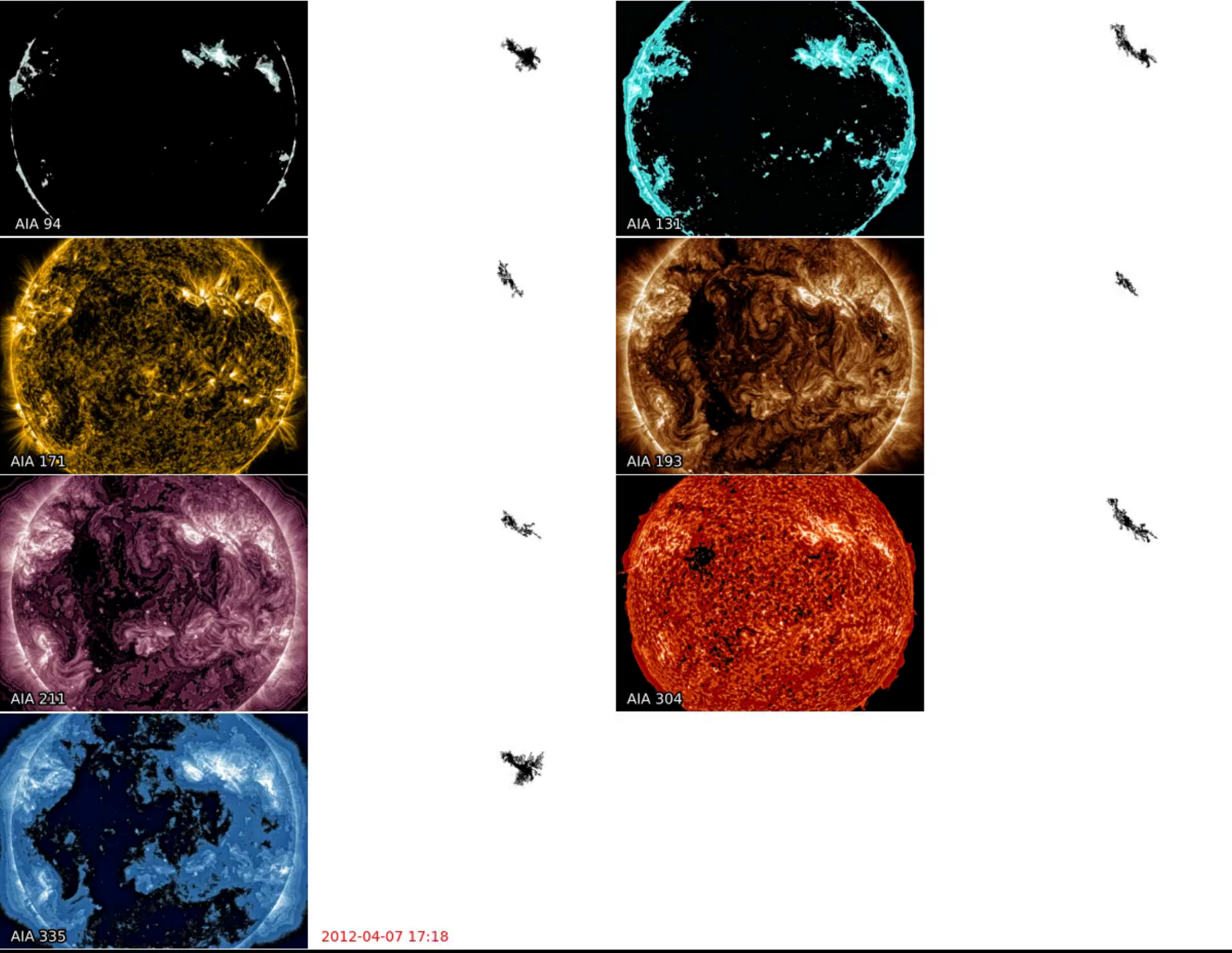}
        \caption{\texttt{ALMANAC} index 10.}
    \end{subfigure}

    \vspace{0.5cm}

    \begin{subfigure}{0.7\textwidth}
        \centering
        \includegraphics[width=\linewidth]{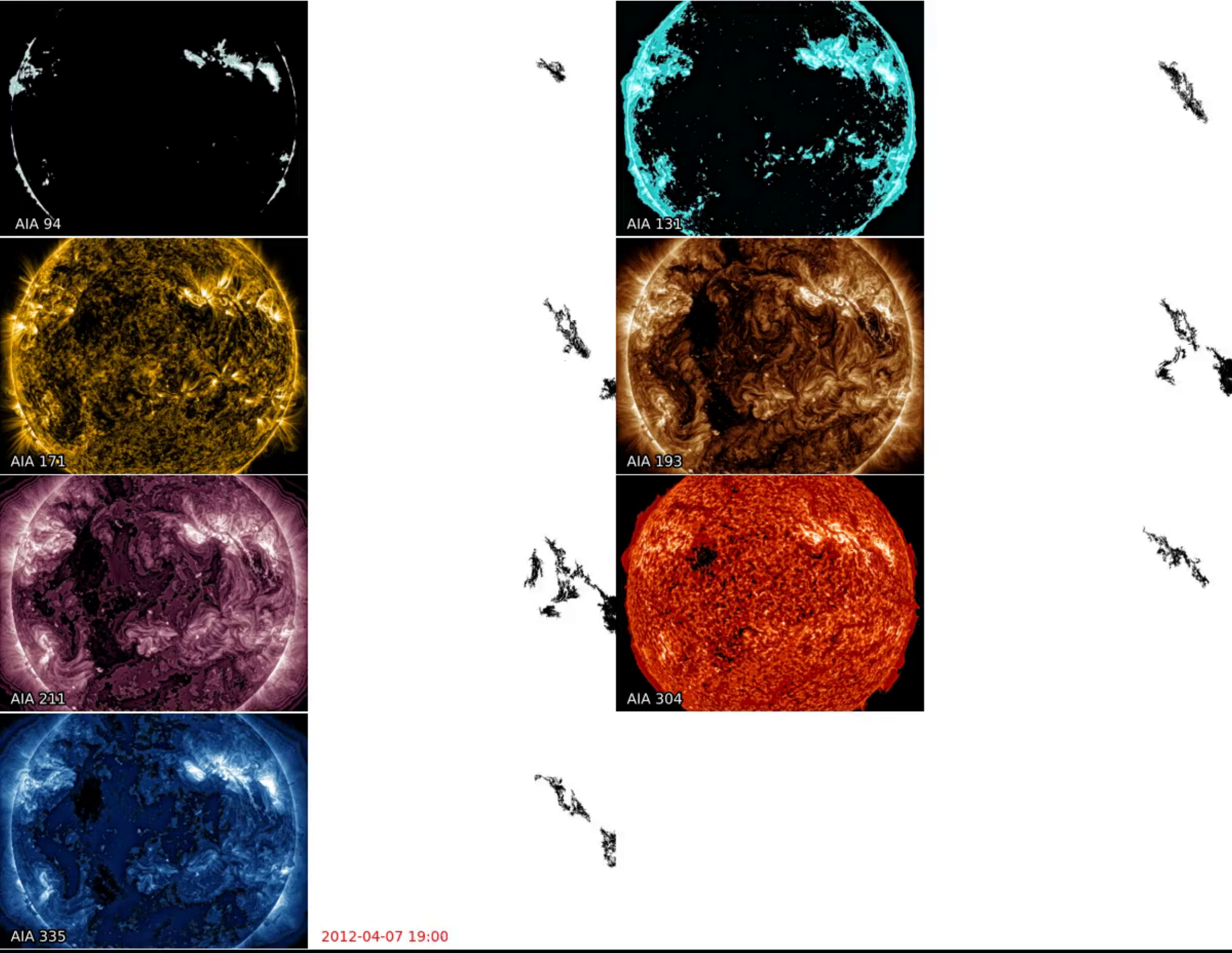}
        \caption{\texttt{ALMANAC} index 11.}
    \end{subfigure}

    \caption{The same as Figure\,\ref{fig:cme2} but for \texttt{ALMANAC} detections 10 and 11 associated with CME (\texttt{CDAW} Index) 5.}
    \label{fig:cme5}
\end{figure}
\begin{figure*}
\centerline{\includegraphics[width=\textwidth]{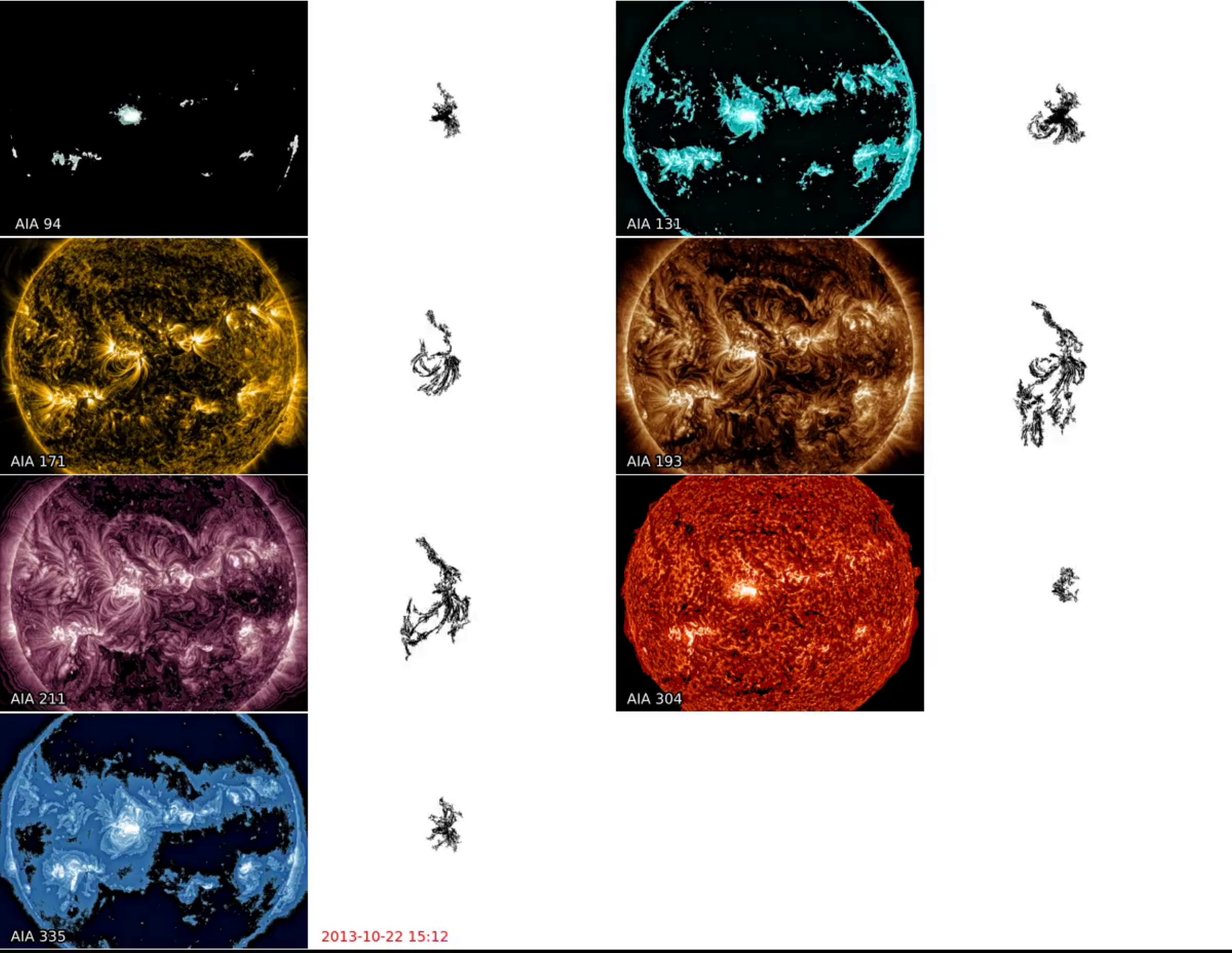}}
\caption{The same as Figure\,\ref{fig:cme2} but for a CME associated with AR\,11875 that was not detected by the original IDL version of \texttt{ALMANAC}. Note that an animated version is only available in the online version.}
\label{fig:cme11875}
\end{figure*}
\begin{table}[ht]
    \centering
    \caption{20 Test Halo-CMEs from \texttt{CDAW} and the corresponding \texttt{ALMANAC} Detections.}
    \begin{tabular}{c c c c c c c c c c}
        \texttt{CDAW}  & Date & Time       & X-ray  & \texttt{CDAW}  & \texttt{CDAW}     & \texttt{ALMANAC} & \texttt{ALMANAC} & \texttt{ALMANAC}  & Eruption \\
        Index &      & (LASCO/C2) & (\texttt{CDAW}) & Onset & Location & Index   & Onset   & Location & Type \\
        \hline
        1 & 2010-08-14 & 10:12 & C4.4 & 09:38 & N17W52 & 1 & 04:12 & N17W43 & False \\
         &  &  &  &  &  & 2 & 08:48 & N04W64 & CME \& Flare \\
        \hline
        2 & 2011-02-15 & 02:24 & X2.2 & 01:44 & S20W12 & 3 & 00:00 & S11W02 & CME \& Flare \\
        \hline
        3 & 2011-09-06 & 02:24 & M5.3 & 01:35 & N14W07 & 4 & 20:24 & N08W01 & False \\
         &  &  &  &  &  & 5 & 21:06 & N07W02 & False \\
         &  &  &  &  &  & 6 & 00:48 & N08W16 & CME \& Flare \\
        \hline
        4 & 2011-11-26 & 07:12 & C1.2 & 06:39 & N17W49 & 7 & 01:12 & S20W35 & False \\
         &  &  &  &  &  & 8 & 01:12 & N18E14 & False \\
         &  &  &  &  &  & 9 & 06:12 & N05W48 & CME \& Flare \\
        \hline
        5 & 2012-04-07 & 21:15 & --- & 19:25 & S24E168 & 10 & 16:36 & N34W16 & CME \& Flare \\
         &  &  &  &  &  & 11 & 17:42 & N19W40 & CME \& Flare \\
        \hline
        6 & 2012-11-23 & 13:48 & B5.8 & 11:00 & S38E10 & 12 & 11:42 & N08W24 & False \\
         &  &  &  &  &  & 13 & 11:48 & S38E09 & CME \& Flare \\
        \hline
        7 & 2012-11-27 & 02:36 & --- & 00:09 & N13E68 & 14 & 21:06 & S16W27 & Flare Only \\
         &  &  &  &  &  & 15 & 01:36 & N64E58 & CME Only \\
         &  &  &  &  &  & 16 & 01:42 & S15W34 & Flare Only \\
        \hline
        8 & 2013-03-15 & 07:12 & M1.1 & 05:46 & N11E12 & 17 & 05:24 & N13E12 & CME \& Flare \\
         &  &  &  &  &  & 18 & 05:48 & N03E02 & CME \& Flare \\
        \hline
        9 & 2013-04-11 & 07:24 & M6.5 & 06:55 & N09E12 & 19 & 06:24 & N12E21 & CME \& Flare \\
        \hline
        10 & 2013-05-15 & 07:12 & X1.2 & 01:25 & N12E64 & 20 & 01:12 & N43E66 & CME \& Flare \\
         &  &  &  &  &  & 21 & 02:00 & N34E69 & Flare Only \\
         &  &  &  &  &  & 22 & 06:06 & S31W36 & CME \& Flare \\
        \hline
        11 & 2013-07-09 & 15:12 & B7.7 & 14:00 & N19E14 & 23 & 09:12 & S21W02 & False \\
         &  &  &  &  &  & 24 & 13:42 & N20E13 & CME \& Flare \\
        \hline
        12 & 2013-08-20 & 08:12 & C1.1 & 06:16 & S31W18 & 25 & 06:48 & S45W18 & CME \& Flare \\
        \hline
        13 & 2013-09-29 & 22:12 & C1.3 & 21:43 & N17W29 & 26 & 21:12 & N17W35 & CME \& Flare \\
        \hline
        14 & 2014-02-18 & 01:36 & --- & 00:30 & S24E34 & 27 & 19:36 & S13W21 & False \\
         &  &  &  &  &  & 28 & 20:00 & S01E04 & False \\
         &  &  &  &  &  & 29 & 00:30 & S31E36 & CME Only \\
        \hline
        15 & 2014-03-23 & 03:36 & C5.0 & 03:05 & S12E40 & 30 & 01:18 & S07E42 & CME \& Flare  \\
         &  &  &  &  &  & 31 & 01:36 & S04W28 & False \\
        \hline
        16 & 2014-04-01 & 16:48 & --- & 14:00 & S09E12 & 32 & 10:48 & S14E18 & False \\
         &  &  &  &  &  & 33 & 11:54 & N26W09 & False \\
         &  &  &  &  &  & 34 & 14:36 & S01E21 & CME Only \\
         &  &  &  &  &  & 35 & 16:06 & S02E58 & False \\
        \hline
        17 & 2014-04-29 & 23:24 & B9.1 & 22:28 & S12E15 & 36 & 22:12 & S07E20 & CME \& Flare \\
        \hline
        18 & 2014-08-15 & 17:48 & --- & 16:16 & S10W05 & 37 & 13:06 & S30W28 & False \\
         &  &  &  &  &  & 38 & 16:18 & S20W12 & CME Only \\
         &  &  &  &  &  & 39 & 16:24 & N06E10 & False \\
        \hline
        19 & 2014-08-22 & 11:12 & C2.2 & 10:13 & N12E01 & 40 & 08:36 & N18E65 & Flare Only \\
         &  &  &  &  &  & 41 & 09:42 & N05W07 & CME \& Flare \\
        \hline
        20 & 2014-12-21 & 12:12 & M1.0 & 11:24 & S14W25 & 42 & 06:48 & S25W45 & CME Only \\
         &  &  &  &  &  & 43 & 10:18 & S12W25 & CME \& Flare \\
        \hline        
    \end{tabular}
    \label{tab:cmes}
\end{table}
\begin{table}[ht]
    \centering
    \caption{\texttt{CDAW} and \texttt{ALMANAC} indexes with absolute time and total spatial differences for CME events.}
    \begin{tabular}{c c c c}
        \texttt{CDAW} Index & \texttt{ALMANAC} Index & $\Delta t$ (min) & $\Delta_{\rm spatial}$ (°) \\
        \hline
        1  & 2  & 50  & 17.69 \\
        2  & 3  & 104 & 13.45 \\
        3  & 6  & 47  & 10.81 \\
        4  & 9  & 27  & 12.04  \\
        6  & 13 & 48  & 1.00 \\
        7  & 15 & 87  & 51.97 \\
        8  & 17 & 22  & 2.00  \\
        9  & 19 & 31  & 9.49  \\
        10 & 20 & 13  & 31.06  \\
        11 & 24 & 18  & 1.41  \\
        12 & 25 & 32  & 14.00 \\
        13 & 26 & 31  & 6.00  \\
        14 & 29 & 0   & 7.28 \\
        15 & 30 & 107 & 5.39 \\
        16 & 34 & 36  & 12.04  \\
        17 & 36 & 16  & 7.07  \\
        18 & 38 & 2   & 12.21 \\
        19 & 41 & 31  & 9.21 \\
        20 & 43 & 66 & 2.00 \\
        \hline
        Mean &  & 40.4 & 11.9 \\
        Std Dev &  & 30.2 & 11.6 \\
    \end{tabular}
    \label{tab:cdaw_almanac}
\end{table}

\subsection{Computational Benchmarks}
The original single-channel implementation of \texttt{ALMANAC} typically processed 8 hours of a single channel of AIA data (at a 6 minute cadence) in $\approx40$\,seconds. The updated multi-thermal version presented here reproduces and parallelises the single-channel method within Python such that the initial algorithm for all AIA wavelengths is computed in $\approx 73\pm 19$\,seconds on a standard octa-core desktop CPU, yielding a large increase in processing speed due to six additional channels being computed in that time. However, the total processing time\footnote{Note that this does not include download times as this depends largely on the JSOC server loads and user internet connection, though an effort has been made to optimise the number of concurrent requests for optimal download speeds.} for an observation window is $\approx132\pm30$\,seconds, due to the additional clustering ($\approx31\pm9$\,seconds) and multi-channel movie ($\approx30\pm11$\,seconds) generation that are now part of the processing pipeline. These processing speeds are sufficient to allow for \texttt{ALMANAC} to be ran as a stand-alone continuous early-warning system for space weather events with cadences on the order of a few minutes with `in-the-loop' forecasters being afforded time to assess each output (if any) visually. However, it is worth noting that for the typical once-per-day forecasts on the likelihood of solar activity, these predictions are made at midnight for the next UT day and so these processing speeds are a nice-to-have rather than being mission critical.

Figures\,\ref{fig:cme2} and \ref{fig:cme5} illustrate example outputs from \texttt{ALMANAC} where CMEs have been detected across multiple wavelengths and collated into single detections. The first CME is a well-documented eruptive flare (X-class flare and CME) from active region AR\,11158 (\texttt{CDAW} Index 2 in Table\,\ref{tab:cmes}). In the animated version (online only), it can be seen that the flare occurs first, which is then followed by the CME. The propagation of which can be seen to spread radially from the epicentre across the solar disk. The corresponding \texttt{ALMANAC} masks in AIA\,193 and 211 can be seen to trace this with the whole disk eventually becoming saturated by the detection. The other detection masks remain centralised on the active region (i.e. the flare). Figure\,\ref{fig:cme5} shows the two \texttt{ALMANAC} detections associated with \texttt{CDAW} index 5 in Table\,\ref{tab:cmes}. What is evident here is that there are two distinct CMEs on the near-side, neither of which are indexed in the \texttt{CDAW} catalogue correctly, as outlined by \citet{almanac} for \texttt{ALMANAC} index 11. This was reported as a bifurcation in the proof-of-concept manuscript \citep{almanac}, however upon updating to a multi-thermal analysis it is now clear that these are two distinct eruptions; something that could not be ascertained from the original single-channel analysis. \texttt{ALMANAC} index 10 (Figure\,\ref{fig:cme5}a) highlights this, particularly in AIA\,94 and 131 where material is seen to be ejected in a northerly direction, and is illustrated in more detail in the Appendix.

As with \citet{almanac}, we investigated the same 20 known halo-CMEs from the \texttt{CDAW} catalogue to benchmark against. In the original study these 20 observation windows yielded 32 separate detections, indicating a detection accuracy of 63\%. For the same 20 observation windows, the multi-thermal version presented in this manuscript yields 43 detections (Table\,\ref{tab:cmes}). However, upon further inspection of these on a case-by-case basis, we find that these observation windows contain not only the 20 known CMEs but additional eruptions that are flares, non-Halo events, or are not included in the full \texttt{CDAW} catalogue, which has been shown to miss narrow CMEs \citep{yashiro08}. Furthermore, for some of the longer duration and broader eruptions, we observe a bifurcation of detections similar to CME 5 in \citet{almanac}. When these are accounted for, the accuracy increases marginally from the single channel version to 65\,\% of detections resulting in a space weather event, though the new method allows for subtle distinction between different event classifications due to the multi-channel processing. As with the single-channel version, the multi-thermal approach presented here remains sensitive to instrumental jitter, small transient events, and bulk motion/waves.

As can be seen from Table\,\ref{tab:cdaw_almanac}, the average event origin separation between \texttt{CDAW} and \texttt{ALMANAC} is $40.4\pm30.2$\,min and $11.9\pm11.6^{\circ}$, which is a marginal deviance from the $37.05\pm29.71$\,min and $11.01\pm10.39^{\circ}$ reported by \citet{almanac}. These differences between the multi-thermal and single-channel versions are to be expected as the method now computes origins based on a clustering of detections rather than solely relying on information from a single centroid. It is also worth noting that we would not expect an exact one-to-one match with \texttt{CDAW} due to differences in how the origin is determined. \texttt{CDAW} predominantly relies on flares/X-ray emission for the CME origin while \texttt{ALMANAC} utilises centre-of-mass calculations for EUV running difference calculations. However, when the outlier results (\texttt{CDAW} Indexes 2, 7, 10, 15 and 20) are excluded the differences between \texttt{ALMANAC} and \texttt{CDAW} become $27.9\pm14.9$\,min and $8.7\pm4.8\,^{\circ}$ indicating clear agreement between the methods.

If we now focus upon these outlier events, we note that for the eruption in \texttt{CDAW} index 2 | \texttt{ALMANAC} index 3, the discrepancy with \texttt{CDAW} is caused due to the fact there is a small eruption/ejection that originates at 2011-02-15 00:00 to the east of the active region. The larger eruptive flare (CME \& X2.2 flare) then follows shortly after in the form of a sympathetic eruption, which is not a distinct detection from the initial one when detected by \texttt{ALMANAC} (Figure\,\ref{fig:cme2}). This is why there is good spatial agreement ($9\,^{\circ}$ separation) but a large difference in the onset time between the methods. Similarly, for \texttt{CDAW} index 7 | \texttt{ALMANAC} index 15, the CME is that of a filament eruption with no associated X-ray emission. In these non-flaring instances, the \texttt{CDAW} catalogue observers rely on lower coronal eruptive signatures such as dimming or eruptive prominences. In \texttt{ALMANAC} no obvious signatures are seen above the defined thresholds for this event until later when a dark filament is observed to lift to the north-west of the active region likely attributed to the eruption in \texttt{CDAW} (Supplementary Movie 1). Here, it is possible that \texttt{ALMANAC} thresholds -- despite being more lenient than \citet{almanac} -- are still too stringent to detect the initial phase of this particular eruption, leading to a delayed onset time and latitude for the reported CME origin being $51\,^{\circ}$ higher than in \texttt{CDAW}.

As for \texttt{CDAW} index 10 | \texttt{ALMANAC} index 20 (Supplementary Movie 2), the data is subjected to strong emission from the flare that saturates the CCD to the north of the eruption and causes the centre-of-mass calculation for the origin to be skewed northwards (in the first frame). The \texttt{CDAW} index 15 | \texttt{ALMANAC} index 30 eruption sees localised loop reconfiguration in AIA\,171, which leads to coronal dimming in AIA\,193 and a flare that is seen in all wavelengths (Supplementary Movie 3). When analysing the full event in the \texttt{CDAW} halo-CME catalogue, it is apparent that the coronal dimming event begins at 02:42 in AIA 193 before the flare onset and subsequent CME being detected in LASCO/C2. As such, the early onset of \texttt{ALMANAC} here is due to the reconnection event that leads to the destabilisation of the active region and subsequent eruptive flare being captured in full by the multi-thermal detection. Similarly, the temporal discrepancy seen for \texttt{CDAW} index 20 | \texttt{ALMANAC} index 43 arises due to \texttt{ALMANAC} detecting an earlier CME that is reported in the full catalogue as a non-halo at 10:48 UTC shortly before the M1.0 flare and halo CME.

Unfortunately, the clustering of detections performed in this updated algorithm does not fully eliminate the duplication of events the original method suffered with. For example, \texttt{ALMANAC} indexes 21 and 40 are flare bifurcations from the \texttt{ALMANAC} index 20 and 41 eruptive flare detections, respectively. Furthermore, \texttt{ALMANAC} indexes 27 and 28 are not only a  bifurcation, but they are also the result of `bad' EUV frames that were not successfully eliminated during preprocessing checks, which cause large areas in the ratio frames to become saturated and results in Boolean masks triggering the threshold criteria for possible detections.

Another example is with \texttt{ALMANAC} indexes 17 and 18. Here, both detection groups capture the eruptive flare, however, shortly afterwards the AIA data experiences a blackout and causes later frames in the Boolean masks to become saturated. As such, index 17 captures the full eruption across all AIA wavelengths and index 18 is a repeat of this in all channels bar 94 and 304.

Furthermore, an additional test is performed based on the findings of \citet{aslam24} where one of thirty-eight CMEs examined could not be detected with the single-channel \texttt{ALMANAC}. Upon recomputing the AIA data for the CME associated with AR\,11875 with the multi-thermal approach adopted here, a detection originates at 13:00\,UTC in AIA\,171 (Figure\,\ref{fig:cme11875}) before a small-scale eruption occurs at 13:30 that is detected in 94, 131, 171, 193, and 211. This micro-flare event even precedes the photospheric magnetic winding signature reported at 13:48 by \citet{aslam24} with the CME originating between 15:00 and 15:06 in all wavelengths.

\subsubsection{Benchmark Summary}
The single-channel implementation of \texttt{ALMANAC} provides a strong foundation for real-time detection of CMEs using AIA data. However, the method is not perfect and has three main issues. The first is that some detections suffer from a bifurcation whereby clusters of pixels are grouped independently for the same event due to spatial proximities being outside the specified thresholds. Secondly, smaller CMEs and/or events with ambiguous low-coronal signatures are difficult to distinguish from topology changes in the low corona such as coronal loop evolution, mass flows, brightenings, and plumes. The third challenge with the single-channel method stems from data gaps and/or instrumental jitter. In these scenarios, the entire solar disk becomes saturated in the running difference data-cubes, and as such lead to false positive detections in the method.

As a result, while the single-channel method is able to quickly process data for an 8-hour window, it requires manual intervention to identify meaningful detections for space weather alongside verification with an external catalogue such as \texttt{CDAW}. The multi-thermal implementation of \texttt{ALMANAC} presented here significantly improves processing capability by parallelising the single-channel algorithm and extending it to all AIA wavelengths, while remaining fast enough for near-real-time space weather monitoring. The multi-thermal analysis improves event interpretation by distinguishing separate eruptions and revealing subtleties that are ambiguous in the single-channel method despite the overall detection accuracy increasing only modestly from 63\% to 65\%. Comparisons with the \texttt{CDAW} catalogue show generally good spatial and temporal agreement (especially when outliers are discarded), with larger discrepancies often attributable to sympathetic eruptions, filament eruptions without strong X-ray signatures, flare saturation effects, or \texttt{ALMANAC} detecting precursor activity earlier than \texttt{CDAW}. 

Additionally, the multi-thermal approach vastly minimises the bifurcation of events seen with the single-channel method. For example, in the single-channel approach, it is not uncommon for a flare to cause false detections in unrelated regions due to the exposure time of AIA decreasing to preserve the CCD. This is because during such events the AIA exposure time decreases from 2\,s to 0.1\,s resulting in a reduced signal-to-noise ratio for those observations. Subsequently, the affected observations have different emission profiles compared to surrounding (unaffected) frames in an observation window, which causes large running differences and triggers the detection thresholds of \texttt{ALMANAC}. The clustering outlined in \S\,\ref{sec:cluster} is a crucial component of the multi-thermal approach which allows the method to eliminate unrelated regions when flares are present during positive detections.

Despite these improvements in ambiguous event detection and eliminating spurious detections, the multi-thermal approach still suffers from some duplicated detections and sensitivity to instrumental artefacts, data blackouts, and saturated frames. It is likely that additional examples and perhaps an ML approach will be needed to identify these autonomously so that they can be discarded before they would appear as a detection for an `in-the-loop' forecaster to analyse.

Whilst only marginal gains in accuracy are seen for the original 20 test cases, the primary benefits of this multi-thermal approach compared to the single-channel analysis come in the form of usability and interpretability. For example, an `in-the-loop' forecaster will be able to view the resultant multi-thermal movies (such as Figures\,\ref{fig:cme2} -- \ref{fig:cme11875}) and deduce if the detection is due to heating/cooling, evolving loop dynamics, or whether it is a genuine space weather event. This is not something that is easily realised on some of the more ambiguous events when only observing a single wavelength without the aid of external catalogues.

\subsection{Kurtosis Forecasting}\label{sec:kurt}
\begin{figure*}
\centerline{\includegraphics[width=0.925\textwidth]{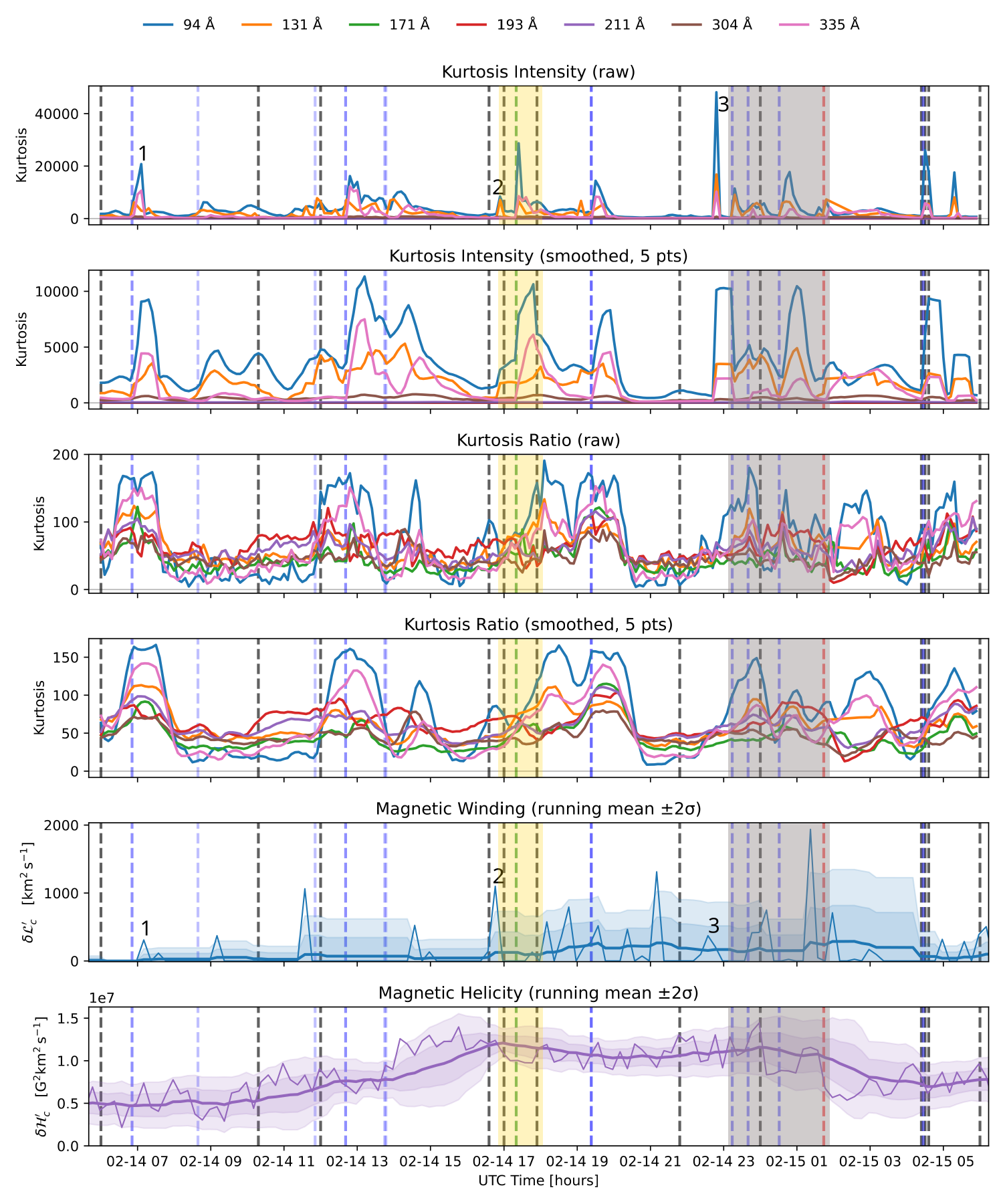}}
\caption{The top four panels show 24-hour plots for kurtosis on the AIA intensity and \texttt{ALMANAC} Ratio datacubes and their 1-hour smoothed time series for NOAA AR\,11158. The bottom two panels shows the corresponding time series for the current-carrying components of the magnetic winding and magnetic helicity. The 1-hour running mean and the 1 and 2-sigma envelopes are shown by the shaded regions. \texttt{ALMANAC} detections are denoted by dashed vertical black lines whilst C, M, and X-class flares are shown as dashed vertical blue, green, and red lines, respectively. The shaded yellow and grey periods indicate strong solar activity after kurtosis and magnetic winding spikes.}
\label{fig:kurt_377}
\end{figure*}
\begin{figure*}
\centerline{\includegraphics[width=\textwidth]{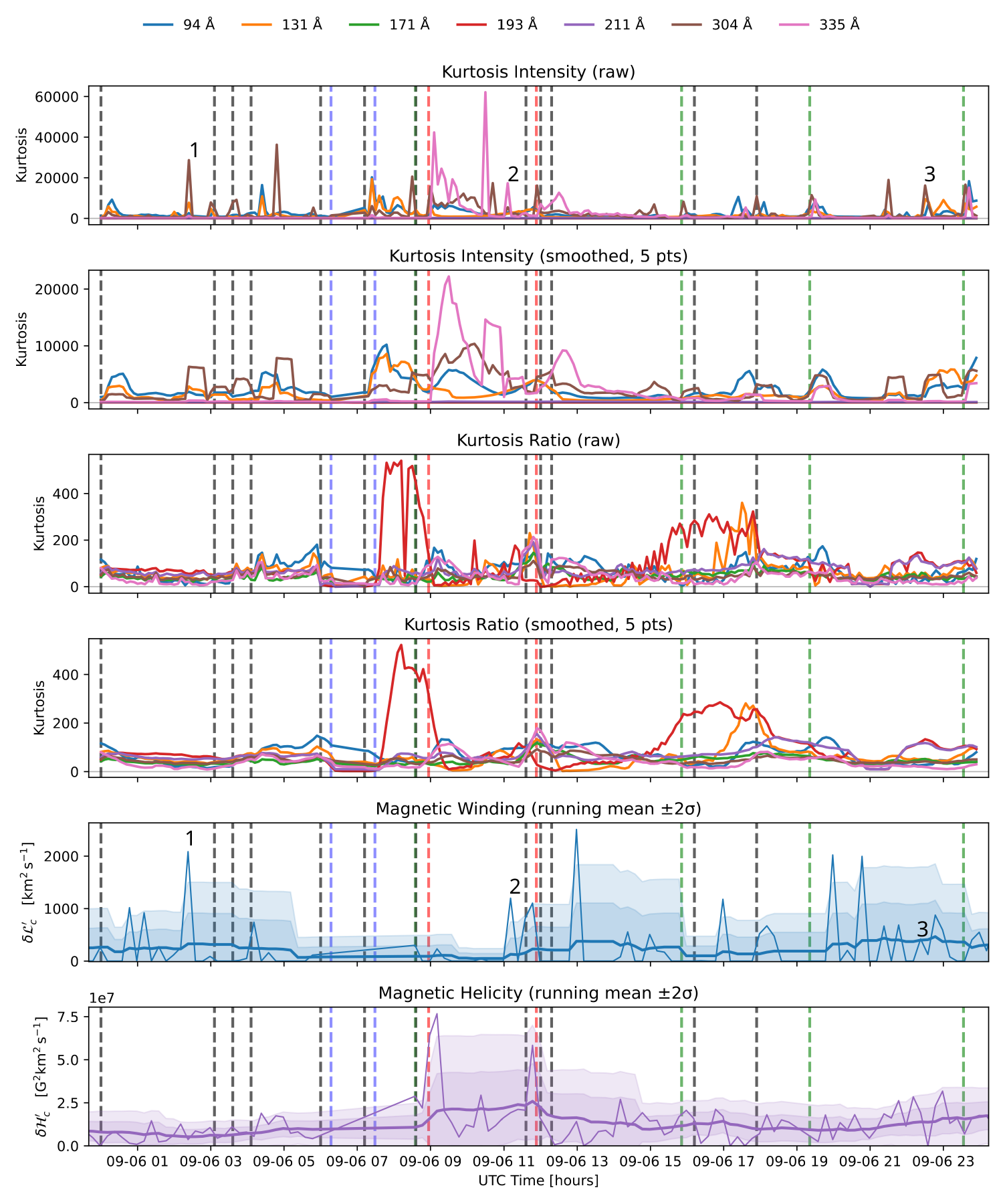}}
\caption{The same as Figure\,\ref{fig:kurt_377} but NOAA AR\,12673}
\label{fig:kurt_7115}
\end{figure*}
\begin{figure*}
\centerline{\includegraphics[width=\textwidth]{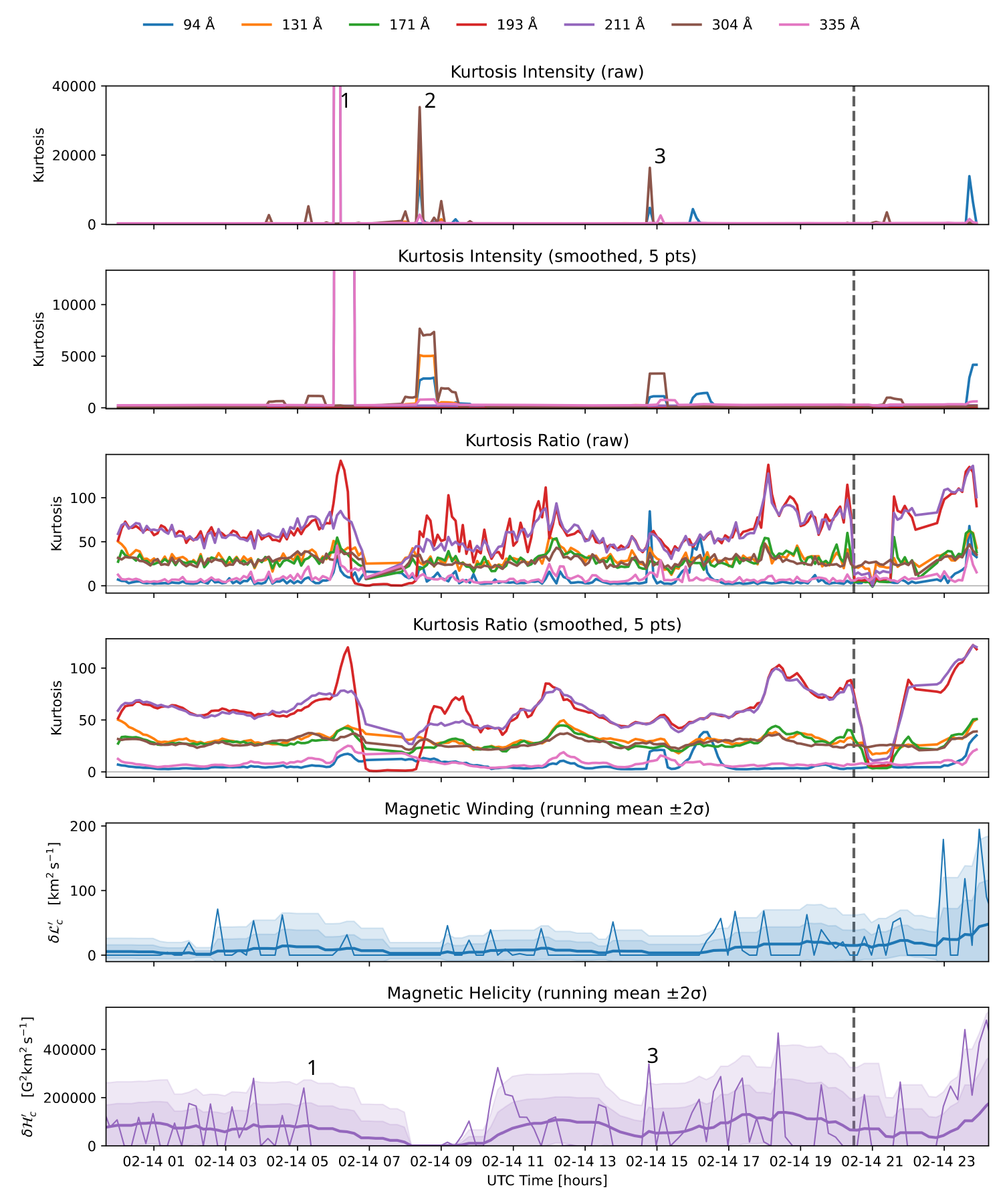}}
\caption{The same as Figure\,\ref{fig:kurt_377} but for NOAA AR\,12699.}
\label{fig:kurt_7237}
\end{figure*}
\begin{figure}[ht]
    \centering

    \begin{subfigure}{0.525\textwidth}
        \centering
        \includegraphics[width=\linewidth]{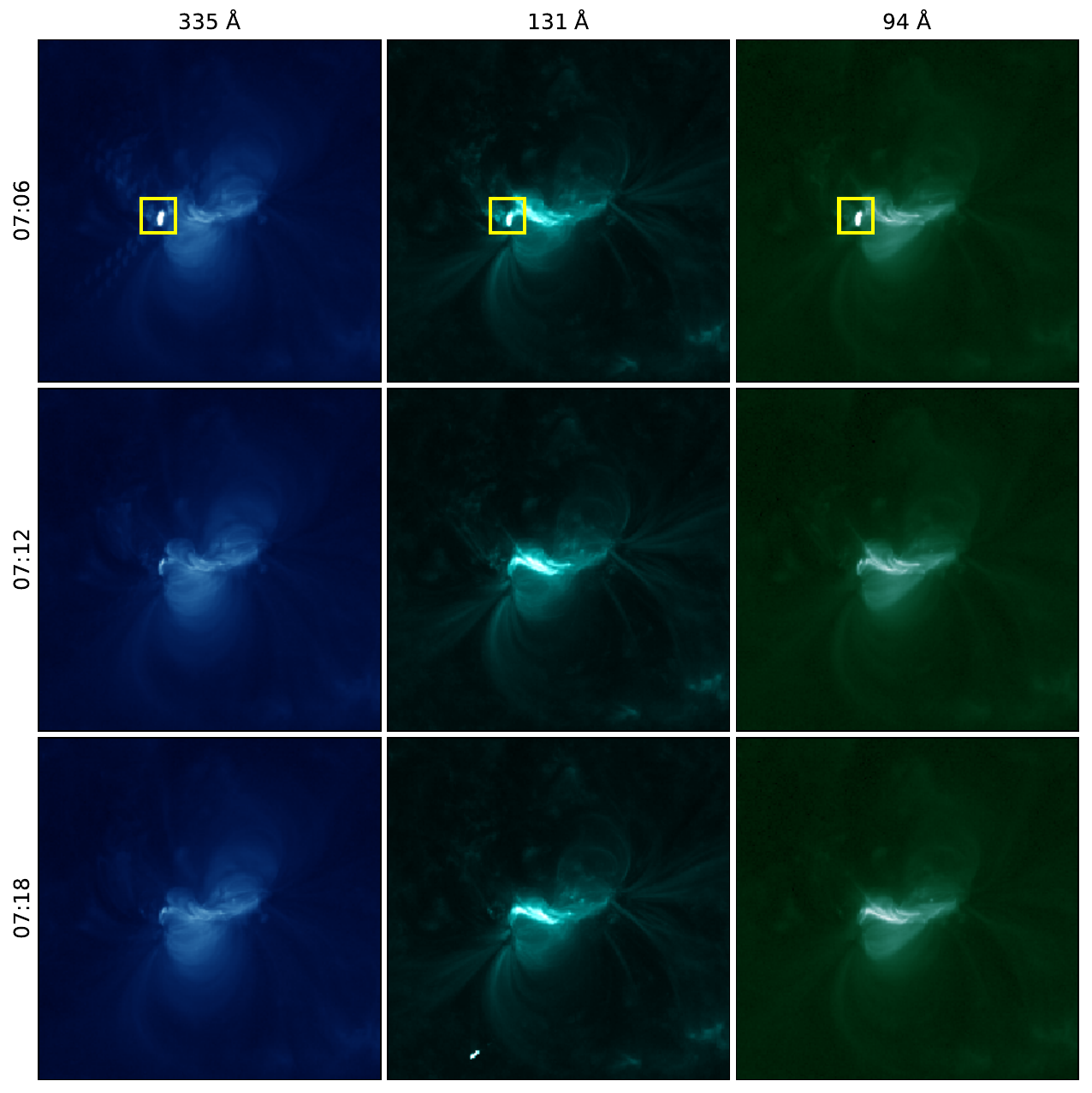}
        \caption{AIA intensity snapshots.}
    \end{subfigure}
    \begin{subfigure}{0.375\textwidth}
        \centering
        \includegraphics[width=\linewidth]{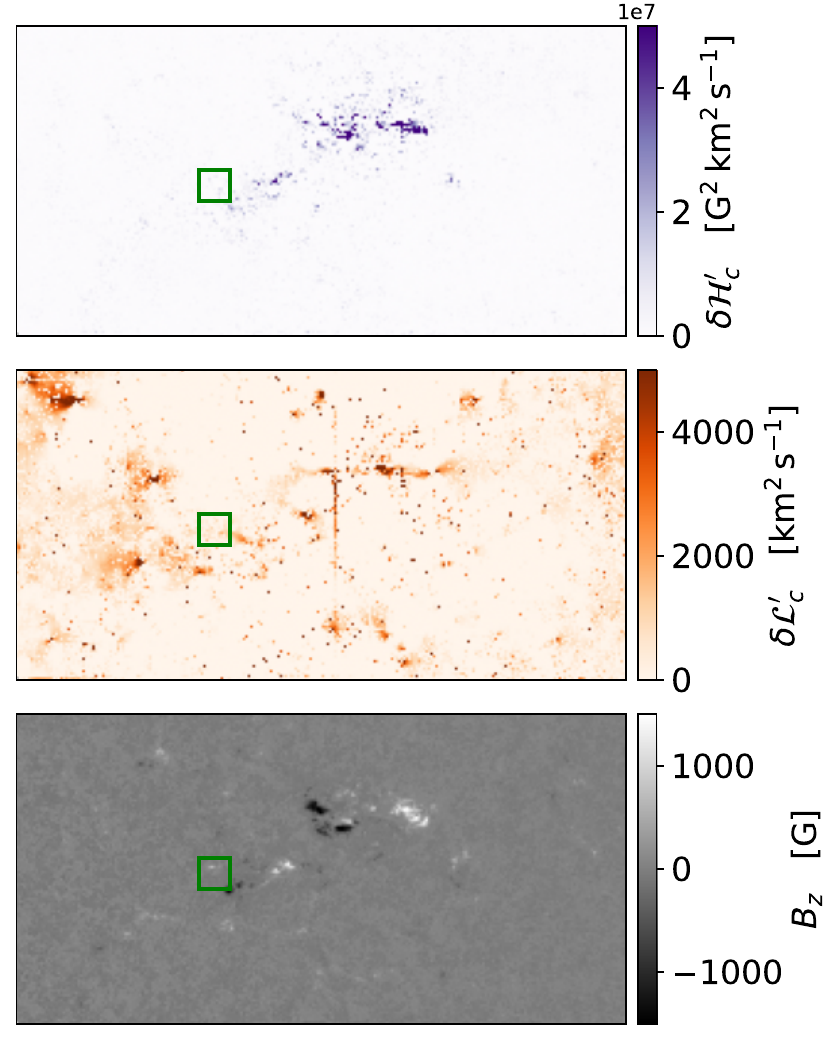}
        \caption{Corresponding $\delta\mathcal{H}_c^{\prime}$, $\delta\mathcal{L}_c^{\prime}$, and $B_z$ at time 07:10.}
    \end{subfigure}

    \caption{NOAA AR\,11158 maps centred around spike 1 in Figure\,\ref{fig:kurt_377}.}
    \label{fig:377_spike1}
\end{figure}
\begin{figure}[ht]
    \centering

    \begin{subfigure}{0.525\textwidth}
        \centering
        \includegraphics[width=\linewidth]{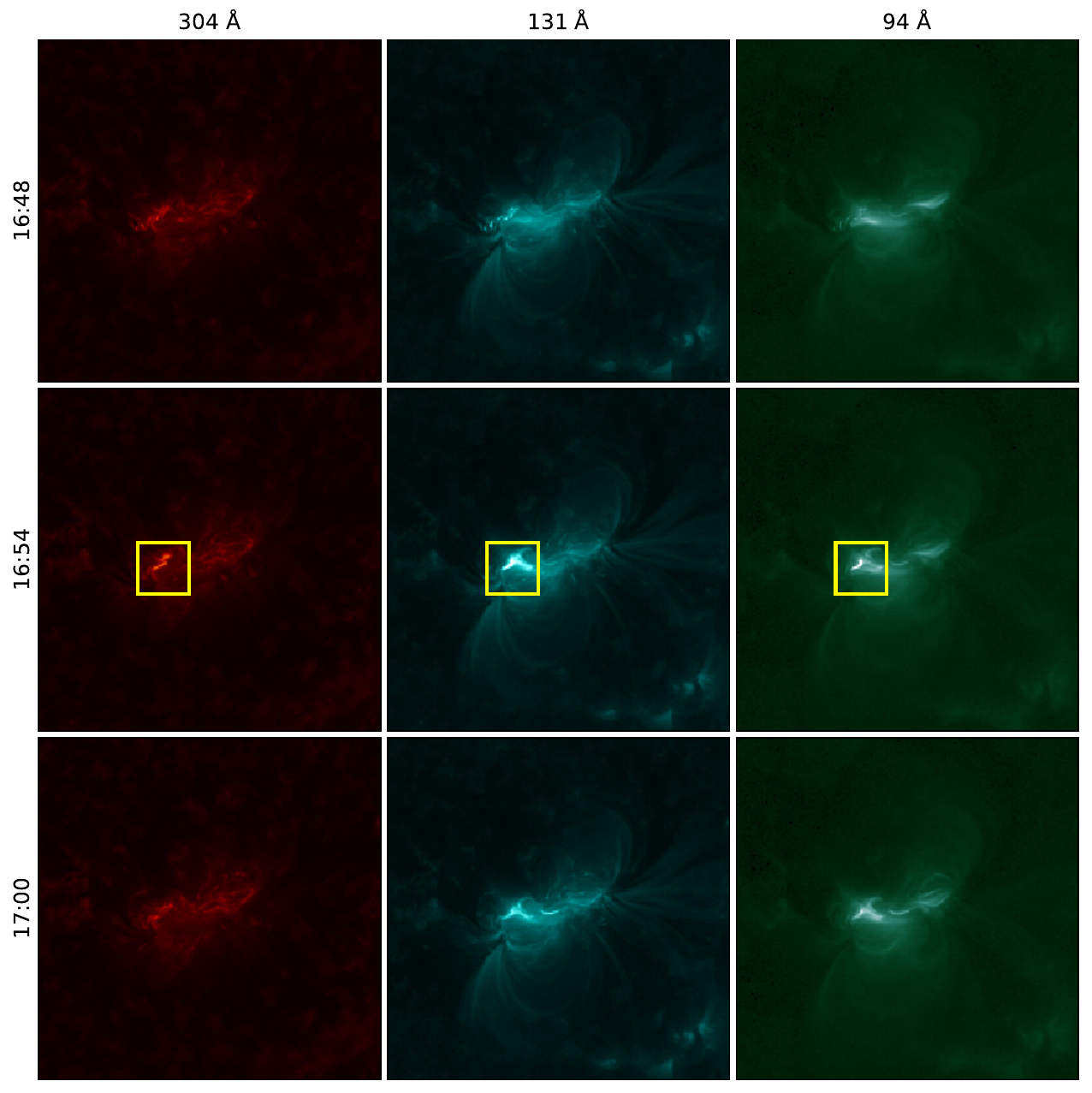}
        \caption{AIA intensity snapshots.}
    \end{subfigure}
    \begin{subfigure}{0.375\textwidth}
        \centering
        \includegraphics[width=\linewidth]{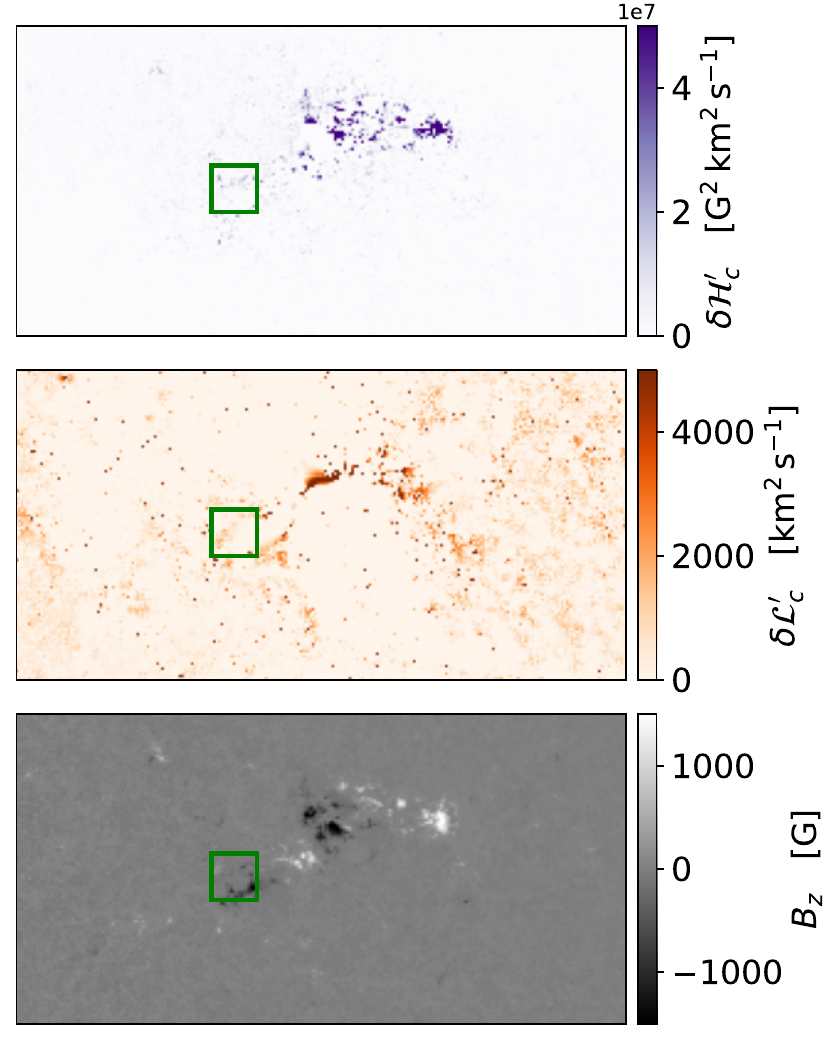}
        \caption{Corresponding $\delta\mathcal{H}_c^{\prime}$, $\delta\mathcal{L}_c^{\prime}$, and $B_z$ at time 16:46.}
    \end{subfigure}

    \caption{NOAA AR\,11158 maps centred around spike 2 in Figure\,\ref{fig:kurt_377}.}
    \label{fig:377_spike2}
\end{figure}
\begin{figure}[ht]
    \centering

    \begin{subfigure}{0.525\textwidth}
        \centering
        \includegraphics[width=\linewidth]{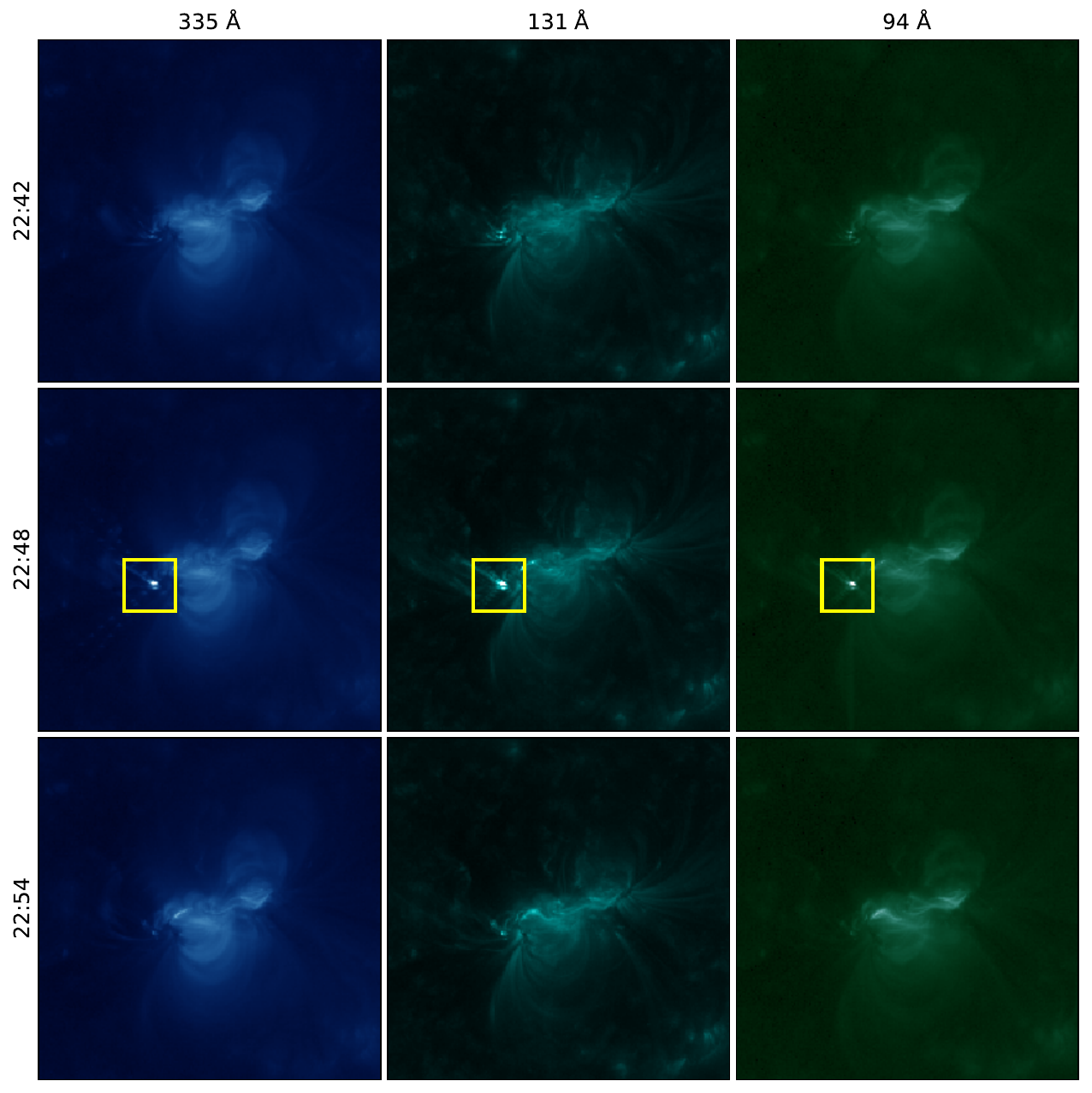}
        \caption{AIA intensity snapshots.}
    \end{subfigure}
    \begin{subfigure}{0.375\textwidth}
        \centering
        \includegraphics[width=\linewidth]{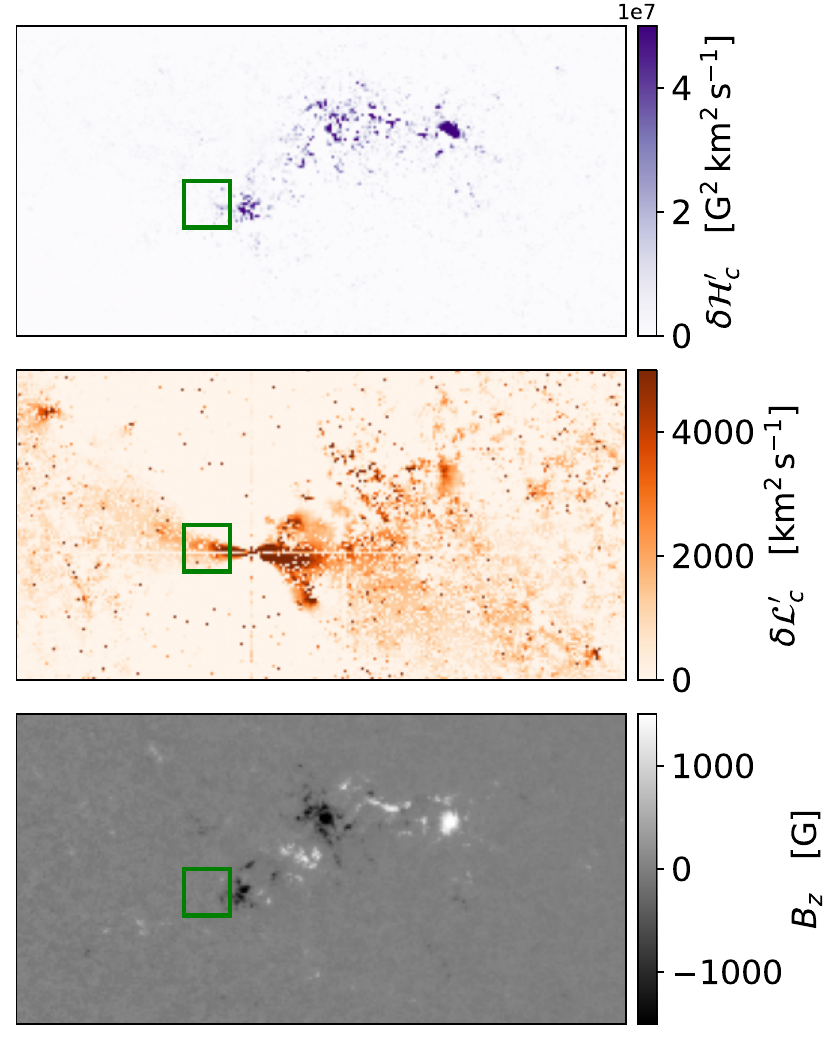}
        \caption{Corresponding $\delta\mathcal{H}_c^{\prime}$, $\delta\mathcal{L}_c^{\prime}$, and $B_z$ at time 22:34.}
    \end{subfigure}

    \caption{NOAA AR\,11158 maps centred around spike 3 in Figure\,\ref{fig:kurt_377}.}
    \label{fig:377_spike3}
\end{figure}
\begin{figure}[ht]
    \centering

    \begin{subfigure}{0.5675\textwidth}
        \centering
        \includegraphics[width=\linewidth]{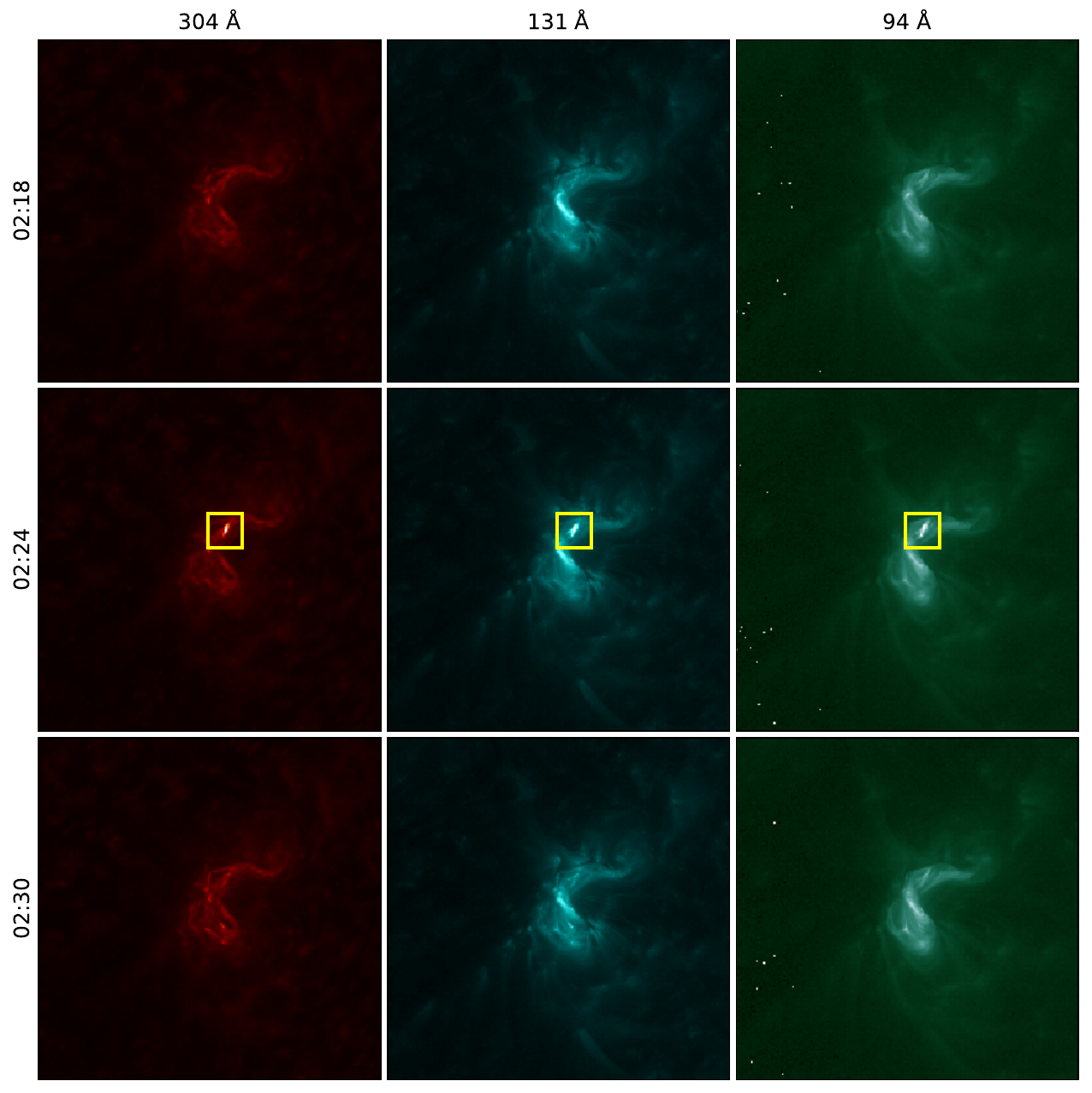}
        \caption{AIA intensity snapshots.}
    \end{subfigure}
    \begin{subfigure}{0.3425\textwidth}
        \centering
        \includegraphics[width=\linewidth]{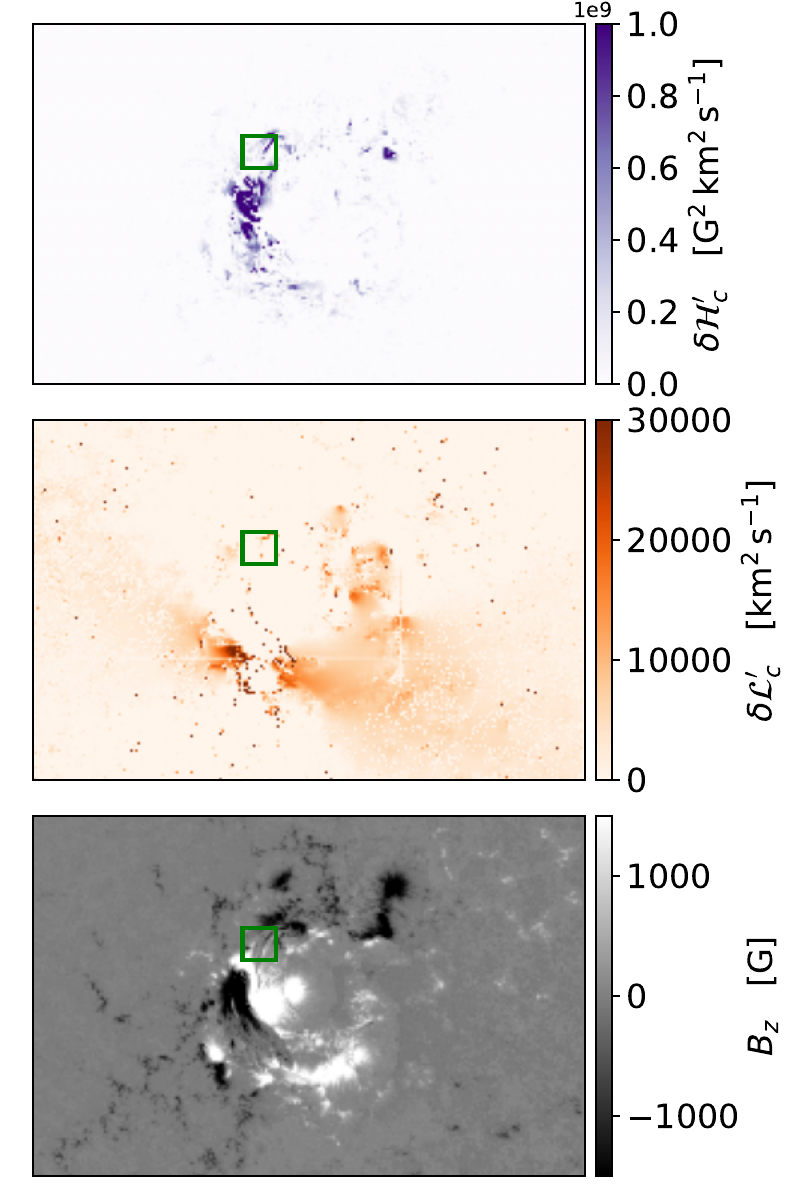}
        \caption{Corresponding $\delta\mathcal{H}_c^{\prime}$, $\delta\mathcal{L}_c^{\prime}$, and $B_z$ at time 02:22.}
    \end{subfigure}

    \caption{NOAA AR\,12673 maps centred around spike 1 in Figure\,\ref{fig:kurt_7115}.}
    \label{fig:7115_spike1}
\end{figure}
\begin{figure}[ht]
    \centering

    \begin{subfigure}{0.5675\textwidth}
        \centering
        \includegraphics[width=\linewidth]{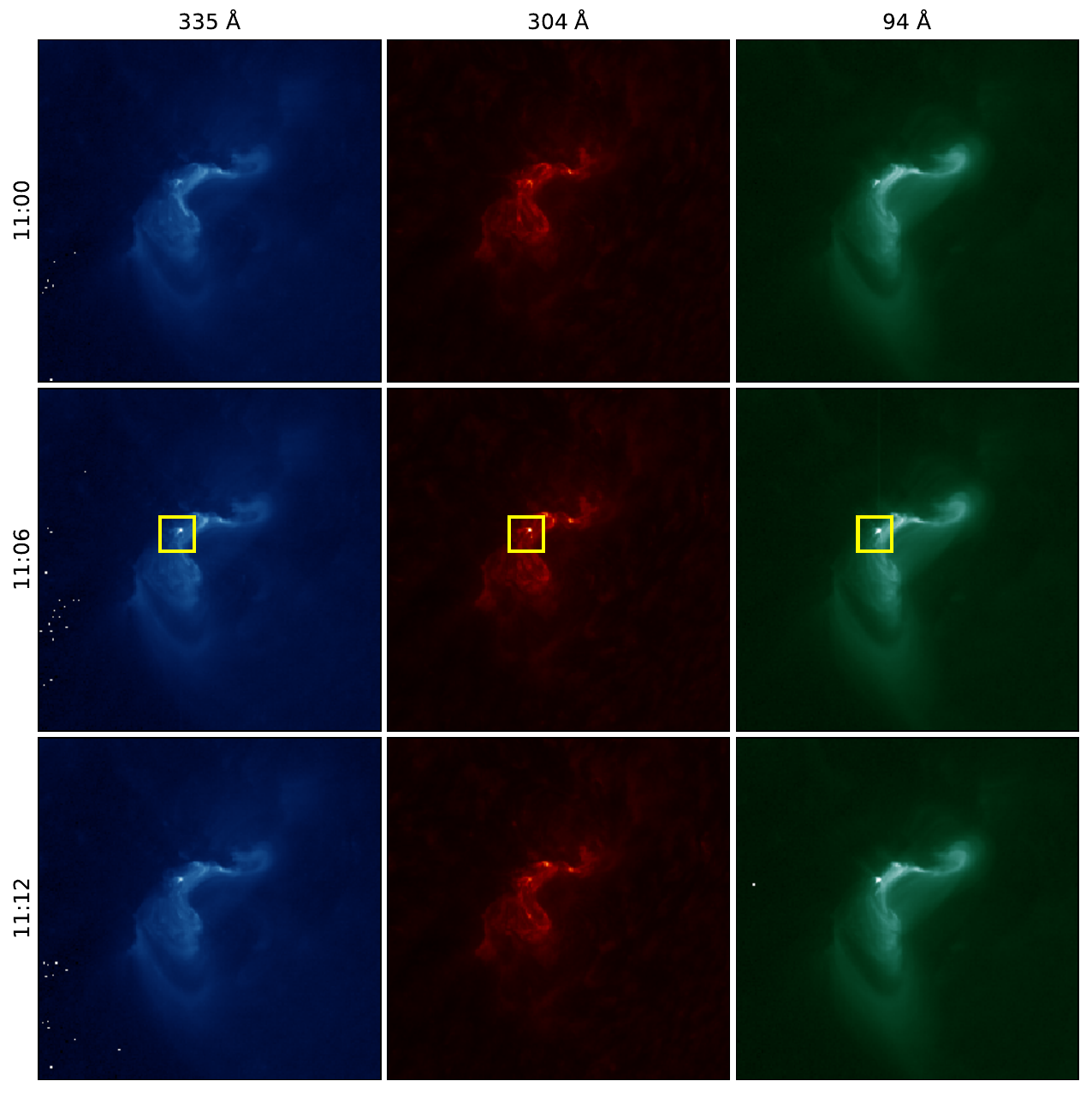}
        \caption{AIA intensity snapshots.}
    \end{subfigure}
    \begin{subfigure}{0.3425\textwidth}
        \centering
        \includegraphics[width=\linewidth]{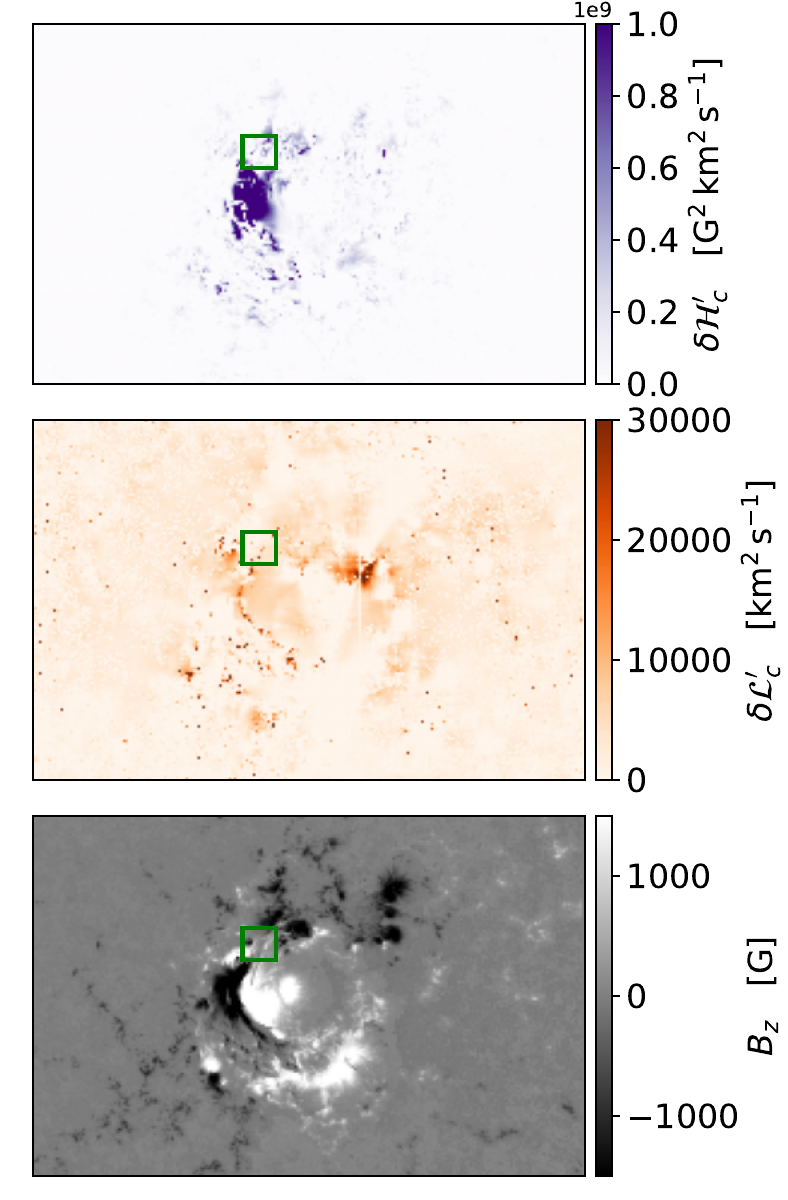}
        \caption{Corresponding $\delta\mathcal{H}_c^{\prime}$, $\delta\mathcal{L}_c^{\prime}$, and $B_z$ at time 11:10.}
    \end{subfigure}

    \caption{NOAA AR\,12673 maps centred around spike 2 in Figure\,\ref{fig:kurt_7115}.}
    \label{fig:7115_spike2}
\end{figure}
\begin{figure}[ht]
    \centering

    \begin{subfigure}{0.5675\textwidth}
        \centering
        \includegraphics[width=\linewidth]{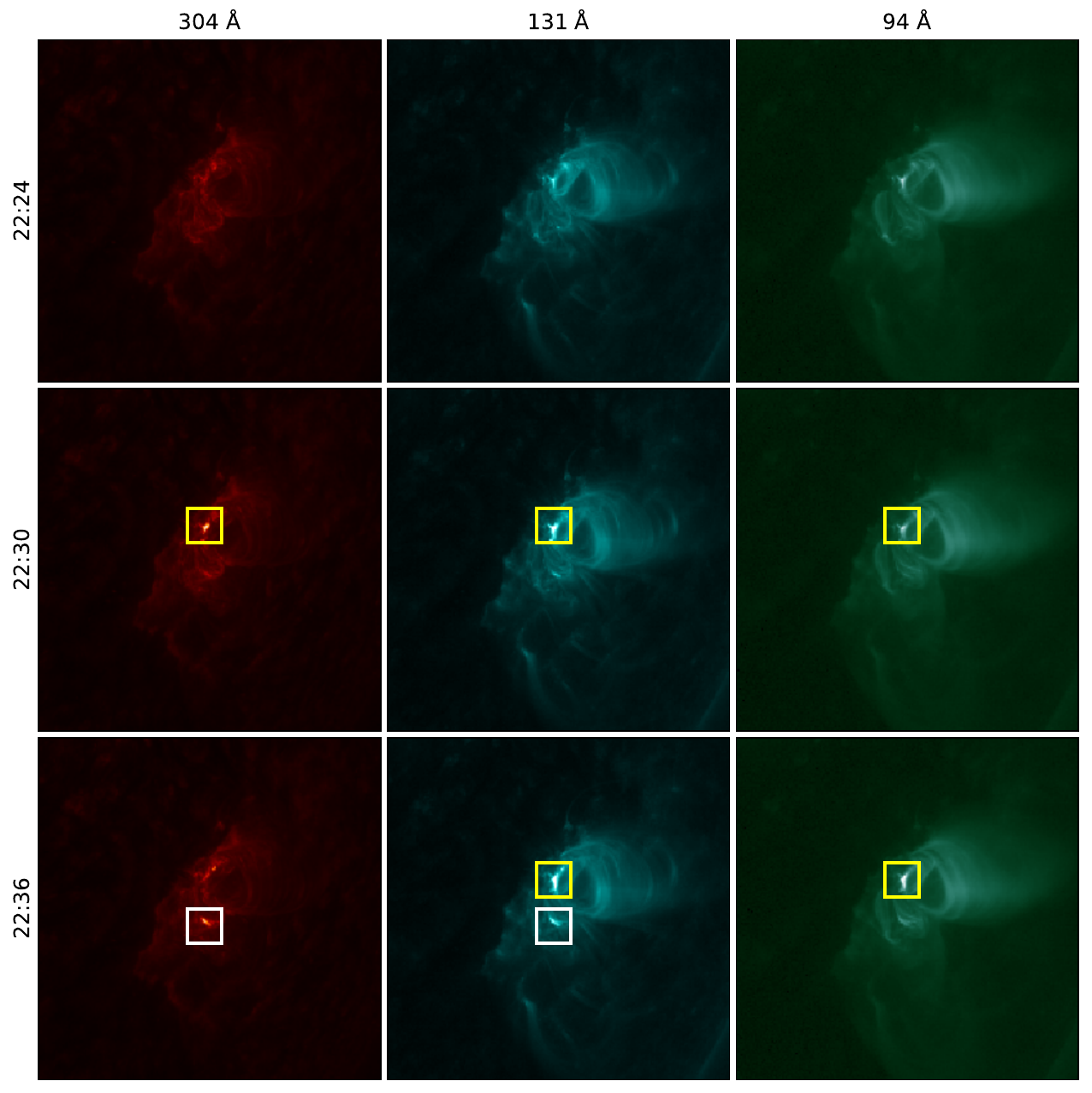}
        \caption{AIA intensity snapshots.}
    \end{subfigure}
    \begin{subfigure}{0.3425\textwidth}
        \centering
        \includegraphics[width=\linewidth]{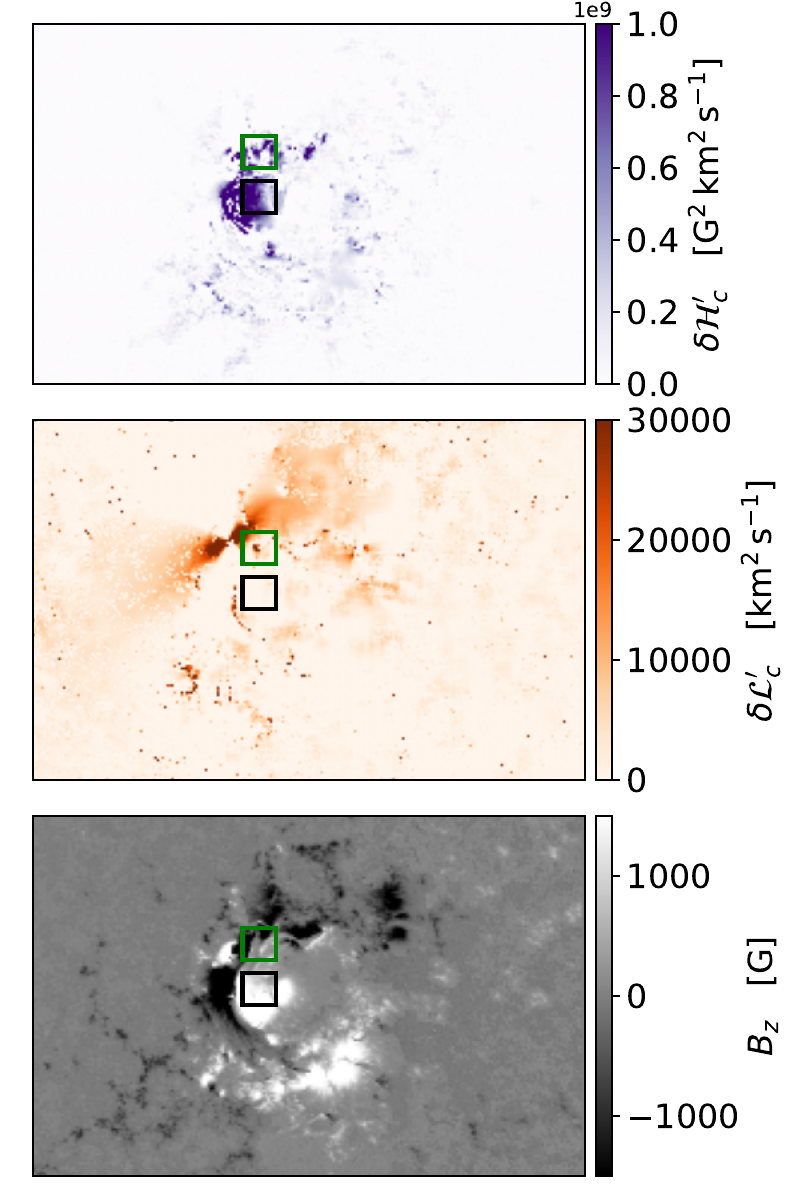}
        \caption{Corresponding $\delta\mathcal{H}_c^{\prime}$, $\delta\mathcal{L}_c^{\prime}$, and $B_z$ at time 22:22.}
    \end{subfigure}

    \caption{NOAA AR\,12673 maps centred around spike 3 in Figure\,\ref{fig:kurt_7115}.}
    \label{fig:7115_spike3}
\end{figure}
\begin{figure}[ht]
    \centering

    \begin{subfigure}[t]{0.475\textwidth}
        \centering
        \adjustbox{valign=t}{
            \includegraphics[width=\linewidth]{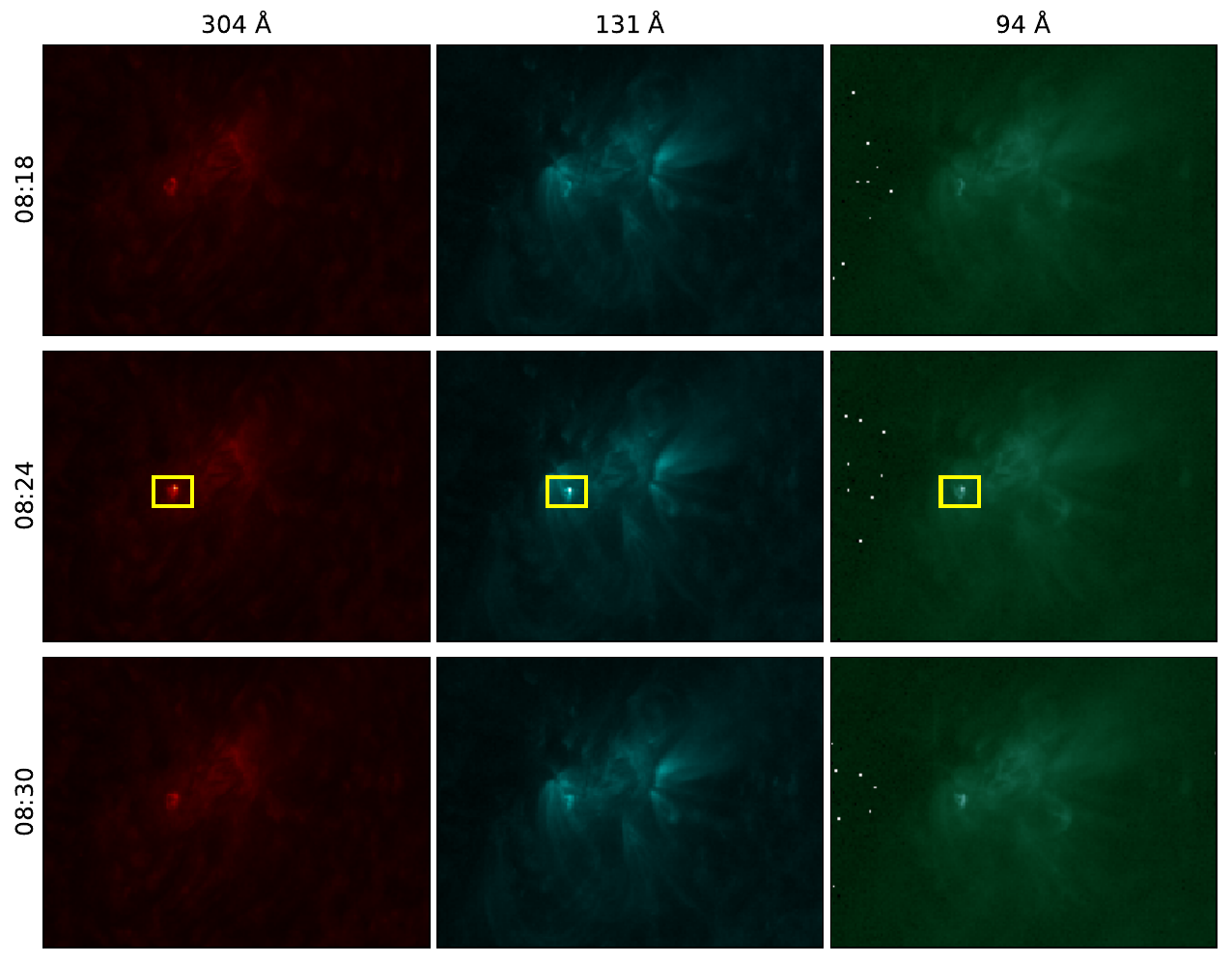}
        }
        \caption{AIA intensity snapshots.}
    \end{subfigure}
    \begin{subfigure}[t]{0.425\textwidth}
        \centering
        \adjustbox{valign=t}{
            \includegraphics[width=\linewidth]{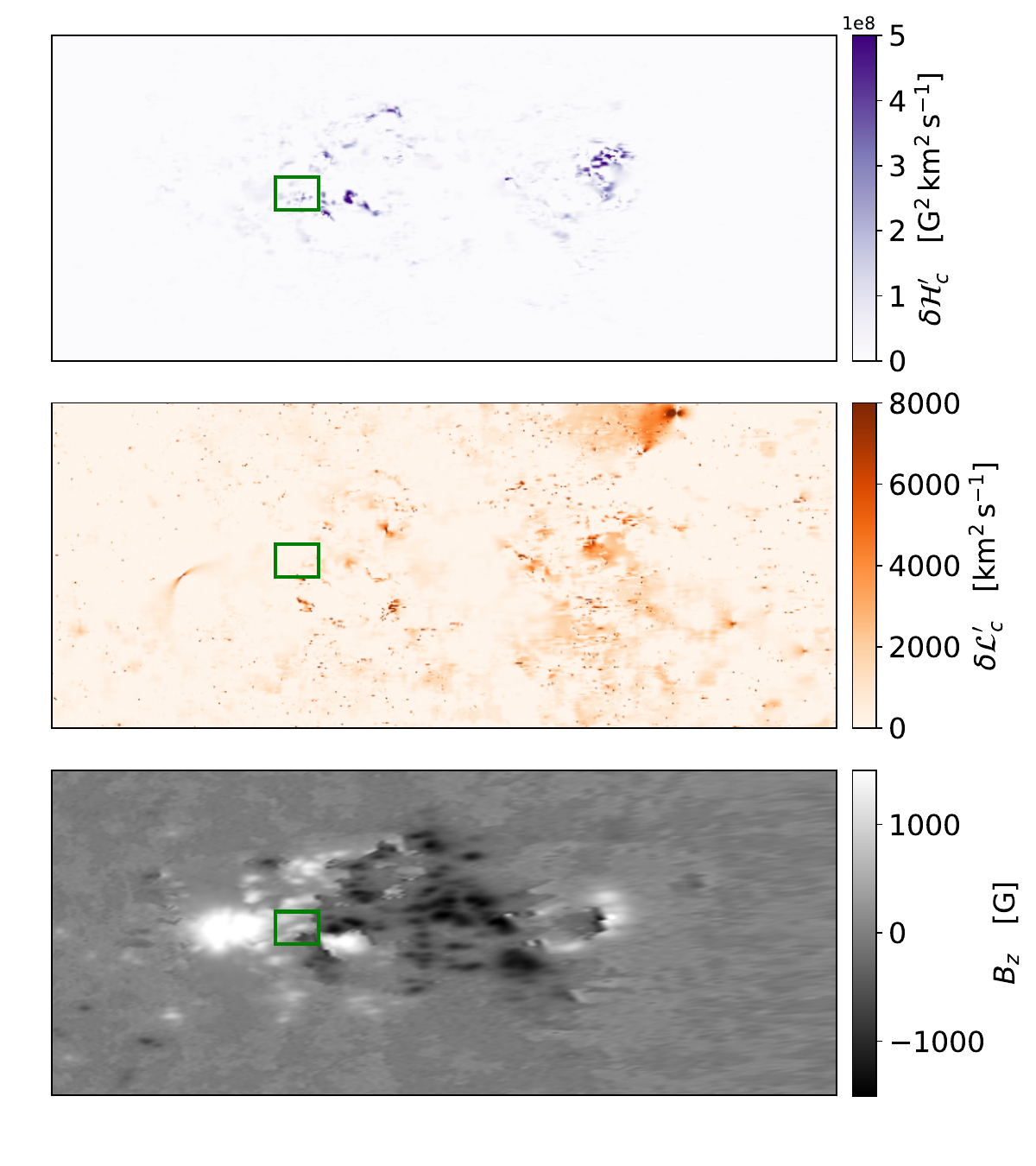}
        }
        \caption{Corresponding $\delta\mathcal{H}_c^{\prime}$, $\delta\mathcal{L}_c^{\prime}$, and $B_z$ at time 08:22.}
    \end{subfigure}

    \caption{NOAA AR\,12699 maps centred around spike 1 in Figure\,\ref{fig:kurt_7237}.}
    \label{fig:7237_spike2}
\end{figure}
\begin{figure}[ht]
    \centering

    \begin{subfigure}{0.475\textwidth}
        \centering
        \includegraphics[width=\linewidth]{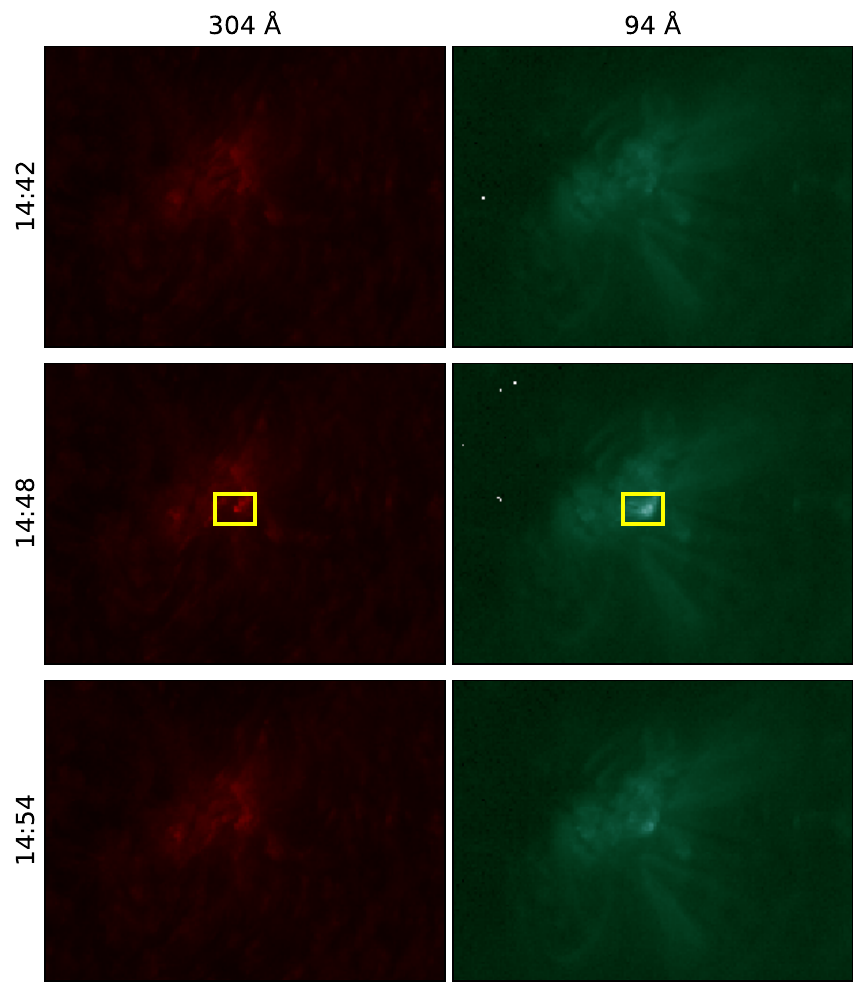}
        \caption{AIA intensity snapshots.}
    \end{subfigure}
    \begin{subfigure}{0.425\textwidth}
        \centering
        \raisebox{0.5cm}{%
            \includegraphics[width=\linewidth]{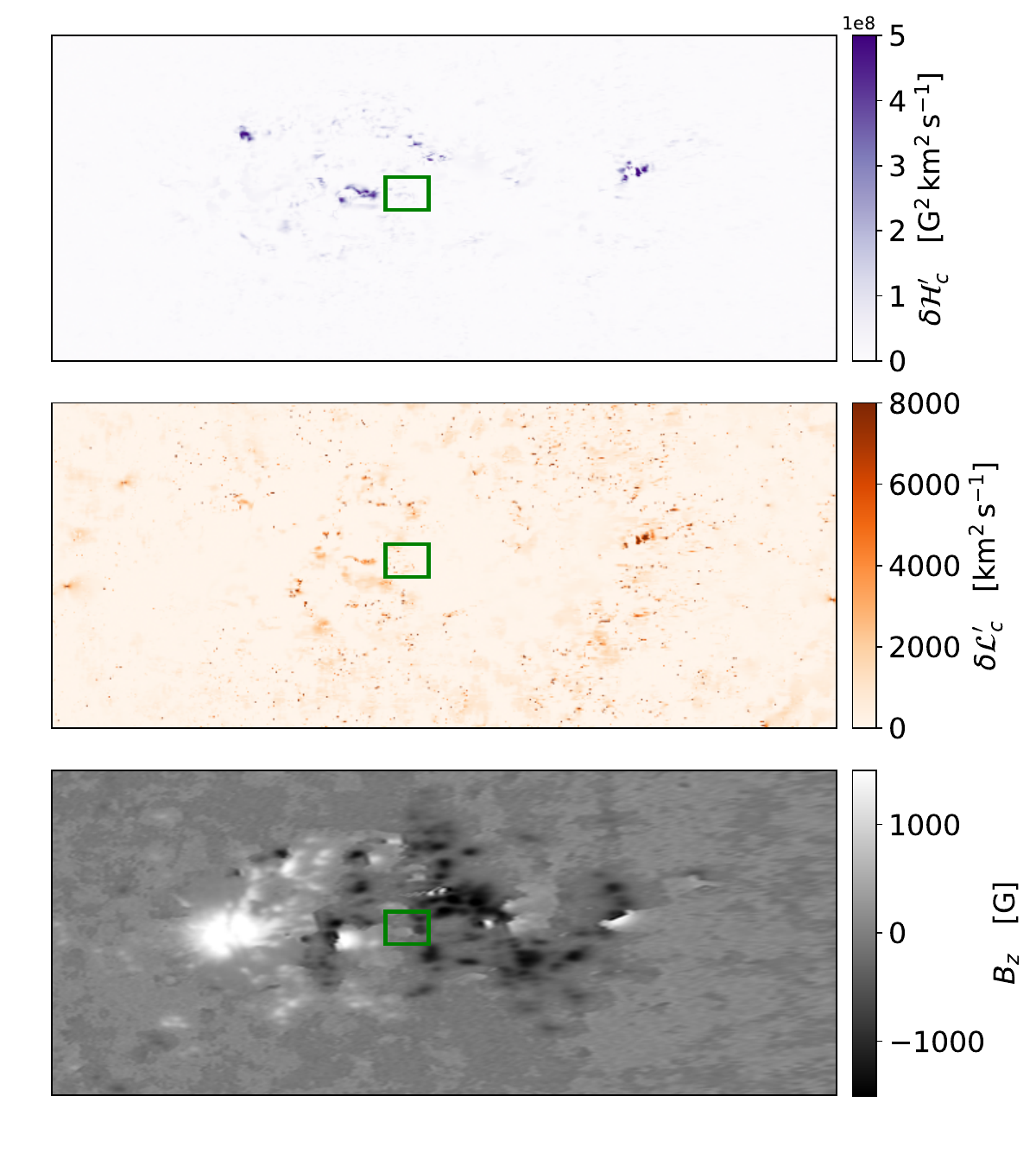}
        }
        
        \caption{Corresponding $\delta\mathcal{H}_c^{\prime}$, $\delta\mathcal{L}_c^{\prime}$, and $B_z$ at time 14:46.}
    \end{subfigure}

    \caption{NOAA AR\,12699 maps centred around spike 3 in Figure\,\ref{fig:kurt_7237}.}
    \label{fig:7237_spike3}
\end{figure}

As mentioned in \S\,\ref{sec:detection}, one of the new metrics calculated in the multi-thermal \texttt{ALMANAC} processing is the kurtosis for the intensity and ratio images of each observation across all wavelengths. The motivation for this is that the kurtosis has shown promising efficacy for solar flare forecasting \citep{williams25,williams26} and is commonly used in the study of system bifurcations in ecology \citep{dakos2019ecosystem}, climate \citep{boers2018early}, the economy \citep{sevim2014developing}, and medicine \citep{chen2012detecting}. Figures\,\ref{fig:kurt_377} and \ref{fig:kurt_7115} show the raw and smoothed running kurtosis series for AIA intensity and ratio outputs from \texttt{ALMANAC} alongside current-carrying components of magnetic winding and magnetic helicity for well-documented eruptive active regions NOAA AR\,11158 and AR\,12673, respectively. Similar plots are shown in Figure\,\ref{fig:kurt_7237} for active region AR\,12699 during a period of no solar activity. In each of these figures, three spikes in the kurtosis intensity have been identified that precede space weather activity and/or align well with spikes in the magnetic winding/helicity time-series. These periods are discussed in detail in the following sections. 

\subsubsection{\texorpdfstring{AR\,11158}{AR 11158}}\label{sec:kurt377}
Spike 1 in Figure\,\ref{fig:kurt_377} can be seen to occur shortly after a C-class flare, with extremal values indicated in spatial maps of the AIA channels 94, 131, and 335 (Figure\,\ref{fig:377_spike1}a) with a spike in the magnetic winding exceeding a $2-\sigma$ envelope during the same period whose spatial maps are shown in Figure\,\ref{fig:377_spike1}b. From the spatial maps here, it is evident that the extremal values in the kurtosis are due to the microflare (yellow boxes) event that occurs to the east of the active region, which is situated next to two negative polarity regions in the magnetogram (green boxes). Within 7 hours of this event, several small flares and a couple of \texttt{ALMANAC} detections occur before a quiet period begins at 14:00 UTC where current-carrying helicity is injected into the region (Figure\,\ref{fig:kurt_377}). 

The second kurtosis time-series spike (Figure\,\ref{fig:kurt_377}) occurs shortly after the first \texttt{ALMANAC} detection of this period, which also precedes a significant spike in the magnetic winding time-series. Here, the sensitivity remains in 94 and 131 but now instead of 335 it is detected in 304, and is again caused by a microflare on the eastern side of the active region, which can be seen in the AIA maps (Figure\,\ref{fig:377_spike2}, yellow boxes). Immediately following this kurtosis spike in Figure\,\ref{fig:kurt_377} is a period of significant solar activity (shaded yellow region) with an M-class flare occurring with \texttt{ALMANAC} detections either side of this event. In the spatial maps, strong magnetic helicity and winding signatures are seen to reside on the opposite side of the coronal structure where the microflare occurs (green boxes). This indicates a potential causal link between the topology input at the photosphere and the microflare that is responsible for the kurtosis spikes and the subsequent solar activity that follows.

As with spike 2, spike 3 in the kurtosis intensity (Figure\,\ref{fig:kurt_377}) shortly follows after a spike in the magnetic winding, though here the spike is below even the $1-\sigma$ threshold due to the larger topology inputs that precede this event. However, it is worth pointing out that the magnitude of spike 3 (in the winding flux time series) is comparable to that of spike 1 in the time-series. Once again, the spike corresponds to a microflare (Figure\,\ref{fig:377_spike3}a; yellow boxes) on the eastern side of the active region in the vicinity of a large and strong magnetic winding input at the photospheric level (Figure\,\ref{fig:377_spike3}b; green boxes). Additionally, there is also a small patch of strong helicity flux in the same location. This is followed by 3 C-class flares and the X2.2 flare and CME (shown in Figure\,\ref{fig:cme2}) and is indicated by the shaded grey period. Interestingly, spike 3 in the AIA\,94 kurtosis intensity is prominently larger than any of the other spikes seen in the time-series for AR\,11158, showing a potential link between pre-eruption kurtosis magnitude and the size of the ensuing solar eruption(s).

Focusing on the kurtosis ratio plots (Figure\,\ref{fig:kurt_377}) we see it is more difficult to make direct comparisons with the topology calculations. Many of the raw ratio inputs for the AIA channels have a tendency to be noise-driven, which is to be expected due to the quantity essentially being a running-difference, but one trend that is easily visualised is that periods of activity tend to have larger smoothed kurtosis ratio values, particularly in AIA 94 and 335, than periods where no activity is logged/detected.

\subsubsection{\texorpdfstring{AR\,12673}{AR 12673}}\label{sec:kurt7115}
We now turn our attention to the three labelled spikes in the AR\,12673 time-series (Figure\,\ref{fig:kurt_7115}). The first spike occurs before a series of three \texttt{ALMANAC} detections in quick succession. Figure \ref{fig:7115_spike1} indicates that the spike is caused by a microflare in the flux rope (yellow boxes) that bridges the two regions of negative polarity at the solar surface. On closer inspection, it appears that the microflare occurs local to a sub-region where there is a mixing of positive and negative magnetic flux at the photosphere (green boxes). Here, the magnetic winding input is dominated by emergence to the south of the SHARP, whilst helicity flux is greater just above this and south of the microflare.

Similar behaviour is seen in spike 2 (Figure\,\ref{fig:7115_spike2}), which occurs immediately before the X9.3 flare and coincident CME. In this instance, a brightening (yellow boxes) that is smaller in area than spike 1, is seen across multiple channels at 11:06 that is once again near the large input of magnetic helicity and opposite polarities of magnetic flux in the photosphere (green boxes). Between the first and second spike highlighted here, the region of helicity input remains consistent whereas the magnetic winding has shifted from the southern footpoint of the flux rope structure seen in the AIA data to the northern one. \citet{williams25} hypothesised that it was the first X-class flare, which caused a disturbance in the flux rope that then leads to the second larger X-class flare. However, the additional information provided by the \texttt{ALMANAC} kurtoses suggests that small-scale reconnection occurred first. This, along with competing inputs of magnetic winding at either end of the flux rope likely disturbed the balance of the magnetic structures modelled by \citet{price19} that led to these sympathetic eruptions from the same magnetic structure. For completeness, we note that here there are a few larger or similar magnitudes spikes in 335 and 304 to spike 2. As these spikes are in single channels at any given time, i.e. they do not have temporal agreement with one another, we interpret that they are likely to be heating/cooling phases from the X2.2 flare that occurs shortly beforehand\footnote{Note that AIA\,335 has several temperature sensitivities from the temperature response function, which makes it a problematic channel to interpret in isolation}. 

Spike 3 corresponds to a small kurtosis intensity, magnetic winding and helicity time-series spikes that do not exceed the running mean envelopes (Figure\,\ref{fig:kurt_7115}). For the magnetic winding, this is due to the fact that two large spikes exceeding $2-\sigma$ occur within close proximity of each other that drastically increase the running mean. Focusing on the spatial maps shown in Figure\,\ref{fig:7115_spike3}, an initial microflare is seen (yellow boxes) which is then immediately followed a secondary microflare (white boxes) in 304 and 131. The same region exhibits a similar shaped structure in 94 though it is much less prominent than the other channels. The photospheric signatures of the first microflare (green boxes) correspond to localised helicity and winding signatures that reside adjacent to the main flux inputs of their respective quantities. The magnetogram shows that this first microflare occurs in a region dominated by negative magnetic flux, though some positive polarity flux does exist here. The second microflare occurs above the region of strongest magnetic helicity input, where no magnetic winding is seen, indicating that this is due to reconnection of near-vertical field lines (black boxes).

The spatial signatures of the magnetic helicity and winding occur in much larger concentrations of strong flux during the period around spike 3 and the other two spikes and is followed by a meaningful event in the form of an M-class flare. However, the time-series spikes (Figure\,\ref{fig:kurt_7115}) do not reflect this. This is due to the fact that the period also sees substantial inputs of potential field, which negates the magnitude of the global $\delta$ quantities as illustrated by equations\,\ref{eq:dL} and \ref{eq:dH}.

As with AR\,11158, the largest spikes seen in the kurtoses for AR\,12673 correspond to the the eruptions with X-class flares. In the kurtosis intensity, the largest spike is seen in AIA\,335 shortly before the X9.3 eruptive flare event. Other large spikes in the kurtosis intensity occur before bursts of activity such as with the three \texttt{ALMANAC} detections following spike 1. In the kurtosis ratio, large sustained values are seen in AIA\,193 before the first X2.2 flare during a period where no topology data exists (and subsequently no spikes before the eruptions).

\subsubsection{\texorpdfstring{AR\,12699}{AR 12699}}\label{sec:kurt7237}
For this example, we focus on a period where there is no reported solar activity in the way of CMEs or flares with a magnitude of C-class and above. For most space weather predictions, it is typical that a flare is only considered meaningful if it is above an M-class in the solar and space weather community (as discussed in \citealp{williams26}). The active region chosen for this is AR\,12699 on 2018-02-14, which had been active with C-class flares and a single M1.17  (2018-02-07 01:45) flare earlier in its life cycle but at this point the last flare (C2.14) occurred at 2018-02-12 01:39. For CMEs, the selected date falls during a quiet period as well, as \texttt{CDAW} reports at least one CME per day during the existence of AR\,12699 bar 2018-02-11, 2018-02-14, 2018-02-16, and 2018-02-18.

The kurtoses shown in Figure\,\ref{fig:kurt_7237} illustrates this lack of activity when compared to that of AR\,11158 (Figure\,\ref{fig:kurt_377}) and AR\,12673 (Figure\,\ref{fig:kurt_7115}). Here, the kurtoses remain `flat' for the duration of the observation window bar a few short duration spikes that typically occur in one or two channels. The distribution for AIA\,304 has a couple of minute spikes that precede an abnormal spike (Spike 1) in AIA 335 for the kurtosis intensity that are only significant due to the surrounding kurtoses being negligible. The abnormal Spike 1 in AIA\,335 is caused by a single bad pixel as the maximum intensity is 3642\,DN\,s$^{-1}$ compared to $\approx80$\,DN\,s$^{-1}$ for the surrounding timestamps.

Spike 2 is interesting from the perspective that it occurs when the $\delta$ quantities for the magnetic topology are dominated by inputs of potential field (zero on y-axes). As is the recurrent theme with these kurtoses spikes when more than one channel is involved, this is caused by a microflare (Figure\,\ref{fig:7237_spike2}; boxed) in the vicinity of small helicity current patches. This spike occurs approximately 3--8 minutes before GOES reports flux of B1.45 and as can be seen in Figure\,\ref{fig:kurt_7237} is followed by another small spike around 09:00 that corresponds to GOES X-ray flux of B1.05. Similar behaviour is seen with spike 3 (Figure\,\ref{fig:7237_spike3}), which temporally matches GOES X-ray flux of B1.54 between 14:45--14:49. In this instance however, there is a statistically meaningful spike in the helicity current that is above a $2-\sigma$ threshold, though it does not seem to be spatially related to the microflare seen in the AIA channels.

The \texttt{ALMANAC} detection at 20:29 is due to a bad frame across 5 AIA channels, which results in the detection thresholds being triggered and a global-level detection is seen. However, whilst no obvious sources for an eruption are seen during this period, we note that a CME is reported by \texttt{CDAW} approximately 4\,hours later at 2018-02-15 00:12 in LASCO/C2. There are two winding spikes and one helicity spike that exceed the $2-\sigma$ envelopes in the hour leading to this detection and notably, a prominent (relative to the time-series) kurtosis spike is seen in AIA\,94 (Figure\,\ref{fig:kurt_7237}).

\subsubsection{Forecasting Efficacy}
Together, these three active regions highlight that kurtosis time-series of AIA data may be useful for forecasting solar eruptions. For example, in AR\,11158, which is a region consisting of bipole pairs, the kurtosis spikes are emphasised most clearly in AIA\,94, which when complemented with spikes in other channels often precede \texttt{ALMANAC} detections and/or singular or a series of flares. AR\,12673 on the other hand, is an emergent region where large influxes of topology lead to magnetic reconnection and solar eruptions. Here, the kurtosis intensity is most consistently spiking before events in AIA\,304, though the largest variations are seen in AIA\,335 between the X2.2 and X9.3 flares. In this instance, the largest spike is $\approx1.5\times$ the magnitude of the largest corresponding spike seen for AR\,11158. Similar behaviour is seen with the kurtosis ratio for AR\,12673 whereby the largest spike is actually a sustained event before the first X-class flare. Unfortunately, this falls during a period where no SHARP data is available and so magnetic topology calculations do not exist to determine if the kurtosis ratio would be complemented by spikes in the topological time-series or if the quantity provides unique insight to potential eruptions that these photospheric quantities cannot.

Furthermore, the largest spikes in the kurtosis intensity for ARs\,11158 and 12673 occur before X-class events and/or to periods prior to an onslaught of eruptions (CMEs and/or flares). This is unlike the magnitude of the spikes seen in helicity and winding, and may provide a tentative link to estimate the potential magnitudes of impending eruptions. As such, these result suggests that inclusion of AIA kurtoses could help flare prediction models \citep{nishizuka2017,nishizuka21,pandey23,bringewald25,williams26} differentiate between M and X-class flares though we do note that spike 1 in AR\,12699 exceeds 100000 on the kurtosis intensity. However, as there is no adjacent activity in the other AIA channels -- such as with the largest spike in AR\,12673 -- it is plausible that a well-trained machine learning model would distinguish this as an outlier when lagged and averaged quantities are considered.

The results from the no activity period of AR\,12699 reveal that the kurtoses are orders of magnitude smaller on the whole compared to the two eruptive regions investigated here. This further reinforces the hypothesis that the kurtosis magnitudes of AIA data may be an important missing link in the forecasting of flares and CMEs that needs further exploration beyond the initial investigation outlined in this manuscript.

\section{Summary}\label{sec:conc}
We present a fully re-engineered Python implementation of the \texttt{ALMANAC} algorithm for automated detection and localisation of CME source regions using multi-wavelength SDO/AIA EUV observations. The pipeline introduces a modular, parallelised framework that enables efficient, high-throughput analysis across all AIA channels, extending the original single-channel approach of \citet{almanac} to a multi-thermal detection suite.

The multi-thermal implementation of \texttt{ALMANAC} introduces three key improvements over the single-channel approach. First, it significantly enhances event interpretation by combining all AIA wavelengths, allowing eruptions, heating/cooling signatures, and evolving coronal structures to be distinguished more clearly. This vastly increases usability on the end-user such as an `in-the-loop' forecaster. Secondly, the inclusion of clustering single-channel detections in space and time greatly reduces bifurcated and spurious detections caused by flare-induced exposure changes and unrelated brightenings. Thirdly, as the method is parallelised across wavelengths, this enables significantly faster processing while remaining suitable for near-real-time space weather forecasting and monitoring.

When the multi-thermal approach is applied to 20 benchmark halo-CME events from the \texttt{CDAW} catalogue, the updated method yields 43 detections, with an effective event association accuracy of 65\%, marginally improving upon previous results while capturing additional eruptions not catalogued in \texttt{CDAW}. The mean discrepancy between \texttt{ALMANAC} and \texttt{CDAW} CME onset times and source locations is $40.4 \pm 30.2$ minutes and $11.9 \pm 11.6^\circ$, respectively, reducing to $27.9 \pm 14.9$ minutes and $8.7 \pm 4.8^\circ$ when excluding outliers. These differences primarily reflect the fundamentally distinct detection paradigms (EUV-based variability versus X-ray/white-light signatures) and the enhanced sensitivity of the multi-channel approach to early-stage or non-flaring eruptions when compared with \texttt{CDAW}.

The kurtosis results presented in this manuscript suggest that the time-series derived from AIA data may provide valuable information for forecasting solar eruptions. Across the eruptive active regions ARs\,11158 and 12673, strong kurtosis spikes frequently align with magnetic topology spikes that precede flares, CMEs, or periods of heightened eruptive activity, with the largest kurtosis spikes often occurring before X-class events. The dominant kurtosis varies between regions and wavelengths, indicating that different eruption mechanisms may produce distinct coronal signatures. In contrast, the relatively quiet AR\,12699 exhibits kurtosis values that are an order of magnitude lower than the eruptive active regions overall, reinforcing the connection between elevated kurtosis and eruptive behaviour. The larger kurtosis magnitudes observed prior to major eruptions also suggest that these quantities may help distinguish between flare classes and improve future flare and CME prediction models when combined with magnetic topology diagnostics and machine learning approaches.

Overall, the multi-thermal \texttt{ALMANAC} framework enhances detection robustness, interpretability, and operational applicability, providing a scalable tool for both retrospective studies and near-real-time space weather monitoring.

\begin{acknowledgements}
T.\,W, C.\,P, and D.\,M acknowledge support from the US Air Force grant FA8655-23-1-7247. D.\,M. acknowledges support from a Leverhulme Trust grant (RPG-2023-182), a Science and Technologies Facilities Council (STFC) grant (ST/Y001672/1) and a Personal Fellowship from the Royal Society of Edinburgh (ID:4282). C.\,P acknowledges support from The UK Science and Technology Funding Council under grant number ST/W00108X/1. This research utilises version 7.1.0 of the \texttt{SunPy} open source software package \citep{sunpy}.
\end{acknowledgements}

\section*{Data Availability Statement}
Our code is published in a GitHub repository: \url{https://github.com/DrTomWilliams/ALMANAC}. The data utilised for this study are provided courtesy of NASA/SDO and the AIA and HMI science teams.

\begin{figure}[ht]
    \centering
    \includegraphics[width=\linewidth]{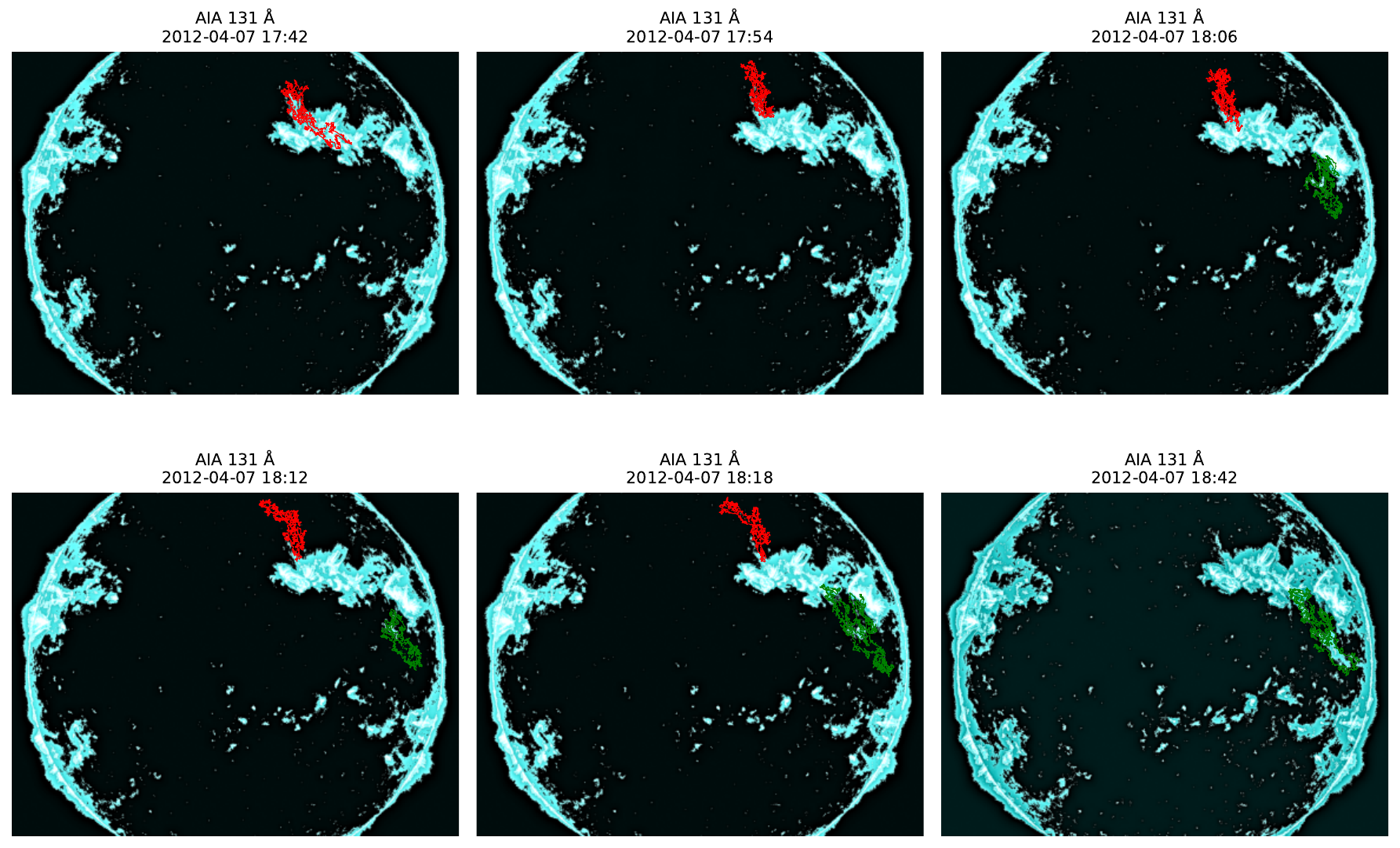}
    \caption{AIA\,131 intensity snapshots that have been enhanced with MGN. The contours of \texttt{ALMANAC} indexes 10 and 11 (Table\,\ref{tab:cmes}) are denoted in red and green, respectively.}
    \label{fig:cme5_131}
\end{figure}

\begin{figure}[ht]
    \centering
    \includegraphics[width=\linewidth]{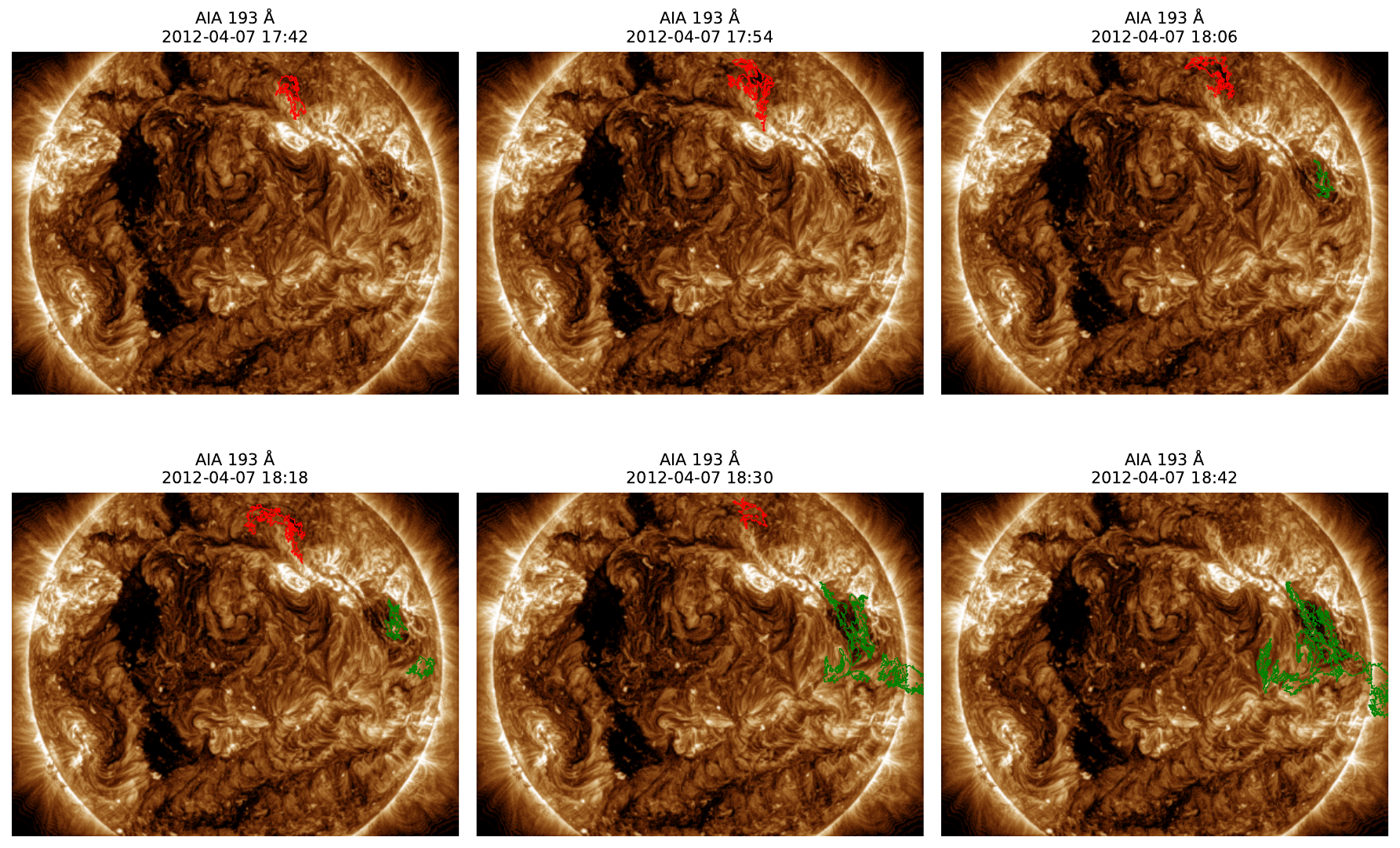}
    \caption{AIA\,193 intensity snapshots that have been enhanced with MGN. The contours of \texttt{ALMANAC} indexes 10 and 11 (Table\,\ref{tab:cmes}) are denoted in red and green, respectively.}
    \label{fig:cme5_193}
\end{figure}

\appendix
Here, we outline in more detail the event of CME\,5 and the two \texttt{ALMANAC} detections (indexes 10 and 11) from Table\,\ref{tab:cmes}. Figures\,\ref{fig:cme5_131} and \ref{fig:cme5_193} outline both \texttt{ALMANAC} indexes 10 (red) and 11 (green) in merged plots for AIA\,131 and AIA\,193, respectively. We see for index 10, the CME originates from the loop arcade above the polarity inversion line for the active region (Figure\,\ref{fig:cme5_131}) at 17:42, which is 24 minutes earlier than the detection for index 11. The animated version (online only) of Figure\,\ref{fig:cme5} indicates that index 10 is an eruptive flare and Table\,\ref{tab:cmes} indicates that it begins approximately an hour earlier than index 11. This point is also illustrated in Figures\,\ref{fig:cme5_131} and \ref{fig:cme5_193} where we can see that index 10 exists in isolation before index 11 is detected and subsequently persisting beyond the lifespan of index 10. Additionally, these combined mask images further illustrate the spatial separation and propagation direction of the two events, further supporting these are two distinct events, and not a bifurcation of the same eruption. However, we do note that it is possible they were triggered by the same underlying mechanism/event, resulting in these two eruptions that have been incorrectly catalogued in \texttt{CDAW} as not only a far-side event but also a non-flaring event.


\bibliography{ref}{}
\bibliographystyle{aasjournalv7}

\end{document}